\documentclass[pdflatex,sn-mathphys-num]{sn-jnl}

\usepackage[utf8]{inputenc}
\usepackage{graphicx}%
\usepackage{multirow}%
\usepackage{hyperref}
\usepackage{amsmath,amssymb,amsfonts}%
\usepackage{amsthm}%
\usepackage{mathrsfs}%
\usepackage[title]{appendix}%

\usepackage{textcomp}%
\usepackage{booktabs}%
\usepackage{algorithm}%
\usepackage{algorithmicx}%
\usepackage{algpseudocode}%
\usepackage{listings}%
\usepackage{bbm}
\usepackage{enumerate}
\usepackage[shortlabels]{enumitem}

\usepackage{xcolor,colortbl}%

\definecolor{LightGreen}{rgb}{0.52, 0.73, 0.4}
\definecolor{LightOrange}{rgb}{0.91, 0.84, 0.42}
\definecolor{LightPurple}{rgb}{0.75, 0.58, 0.89}

\usepackage{tabularx}
\usepackage{booktabs}
\usepackage{pifont}

\usepackage{array}
\newcolumntype{Y}{>{\centering\arraybackslash}X}

\newcommand{\cmark}{\scalebox{1.4}{\ding{51}}} 
\newcommand{\xmark}{\scalebox{1.4}{\ding{55}}} 

\newcolumntype{C}[1]{>{\centering\arraybackslash}m{#1}}  
\newcolumntype{L}[1]{>{\raggedright\arraybackslash}m{#1}} 

\theoremstyle{thmstyleone}%
\theoremstyle{thmstyletwo}%

\theoremstyle{thmstylethree}%

\begin{document}

\title[A Simple and Explainable Model for Park-and-Ride Car Park Occupancy Prediction]{A Simple and Explainable Model for Park-and-Ride Car Park Occupancy Prediction}


\author*[1,2]{\fnm{Andreas} \sur{Kaltenbrunner}}\email{kaltenbrunner@gmail.com}

\author[2]{\fnm{Josep} \sur{Ferrer}}\email{rferrers@gmail.com}

\author[2]{\fnm{David} \sur{Moreno}}\email{david8more@gmail.com}

\author[2]{\fnm{Vicenç} \sur{Gómez}}\email{vicen.gomez@upf.edu}

\affil*[1]{\orgdiv{Internet Interdisciplinary Institute}, \orgname{Universitat Oberta de Catalunya},
  \orgaddress{\street{Rambla del Poblenou 154-156},
  \postcode{08018}
  \city{Barcelona},
  \country{Spain}}}

\affil[2]{\orgdiv{Department of Engineering}, \orgname{Universitat Pompeu Fabra}, \orgaddress{\street{Carrer de T\`anger 122-140}, \postcode{08018} \city{Barcelona}, \country{Spain}}}




\abstract{In a scenario of growing usage of park-and-ride facilities, understanding and predicting car park occupancy is becoming increasingly important. This study presents a model that effectively captures the occupancy patterns of park-and-ride car parks for commuters using truncated normal distributions for vehicle arrival and departure times. The objective is to develop a predictive model with minimal parameters corresponding to commuter behaviour, enabling the estimation of parking saturation and unfulfilled demand.

The proposed model successfully identifies the regular, periodic nature of commuter parking behaviour, where vehicles arrive in the morning and depart in the afternoon. It operates using aggregate data, eliminating the need for individual tracking of arrivals and departures. The model’s predictive and nowcasting capabilities are demonstrated through real-world data from car parks in the Barcelona Metropolitan Area. A simple model extension furthermore enables the prediction of when a car park will reach its occupancy limit and estimates the additional spaces required to accommodate such excess demand.
Thus, beyond forecasting, the model serves as a valuable tool for evaluating interventions, such as expanding parking capacity, to optimise park-and-ride facilities.}




\maketitle

\section{Introduction}
The rapid increase in energy prices and the growing awareness of the need to reduce greenhouse gas emissions have led an increasing number of commuters to shift from private to public transport. Additionally, major urban centres such as London, Barcelona, and Vienna are progressively restricting access to city centres for private vehicles or limiting free-of-charge parking by expanding regulated parking zones. In response to these developments, park-and-ride systems have become a crucial solution, allowing commuters to leave their private vehicles at designated facilities outside city limits and complete their journeys using public transport.


Effective park-and-ride systems require large parking facilities strategically located near major transport hubs, such as train stations. Accurately estimating the required capacity of these parking facilities and expanding them when demand exceeds available space is a critical aspect of urban planning. A thorough understanding of commuter behaviour and the ability to predict parking demand are essential for making informed decisions in this context.



In this work, we use data collected from a set of eight park-and-ride facilities located in the vicinity of Barcelona, Catalonia. These car parks\footnote{Note: throughout this paper, we use the term ``car park'' following British English usage, which corresponds to ``parking lot'' in American English.} primarily serve commuters on working days, who typically park their vehicles in the morning before transferring to public transport for work or study in the city. In the afternoon, they return to retrieve their cars and drive home. 

The observed periodic patterns in car park occupancy are highly regular, making them well-suited for analysis and prediction using simple models. Our goal is to develop models that improve the planning and management of park-and-ride infrastructure. The specific objectives of our modelling approach are as follows:
\begin{itemize}
\item Use as few parameters as possible.
\item Ensure that the model parameters are interpretable and correspond to behavioural metrics.
\item Enable the model to predict when a parking lot will be full.
\item Allow the model to assess the unmet demand for free parking spaces in a car park.
\end{itemize}

To achieve these goals, we will model car arrivals and departures separately, each using a different truncated normal distribution. This approach allows us to use aggregate data without the need to monitor individual arrivals and departures.
Before presenting the proposed model in Section~\ref{sec:model}, we first review relevant literature in Section~\ref{sec:related}. In Section~\ref{sec:dataset}, we describe the dataset used in our study. Section~\ref{sec:results} then analyses the model's ability to fit the observed data and evaluates its performance in both prediction and nowcasting tasks. Finally, we conclude with policy recommendations in Section~\ref{sec:policy} and a discussion of our findings in Section~\ref{sec:discussion}.



\section{Related work}
\label{sec:related}


The literature on smart parking solutions is extensive, spanning multiple disciplines and problem types, as highlighted in the survey by~\citet{survey}. Our work focuses on a specific aspect within this area: modelling car park occupancy. Traditionally, this has been approached primarily as a prediction task (see~\cite{xiao2023parking} for a recent overview). Prediction problems can be classified into two main categories: on-street parking and off-street parking (e.g., garages and dedicated parking lots). Our research specifically addresses the latter.

This section is organised around three main limitations identified in existing studies: limited model interpretability, challenges in handling aggregate data instead of individual vehicle arrivals and departures, and insufficient capacity to account for unmet demand, i.e., reasoning about counterfactual scenarios.

Notable studies that focus specifically on car park occupancy prediction include~\cite{Zheng2015ParkingAP}, \cite{Camero2018EvolutionaryDL}, \cite{zhao2020comparative}, and \cite{awan2020comparative}. In~\cite{Zheng2015ParkingAP}, the authors compared regression trees, support vector regression, and single-layer neural networks for predicting occupancy in sensor-equipped on-street parking in San Francisco (US) and Melbourne (Australia). They concluded that regression trees, using historical occupancy rates and weekdays as input features, provided the best performance with the least computational cost.

Building on this, recurrent neural networks with multiple hidden layers were employed by~\cite{Camero2018EvolutionaryDL} to predict occupancy rates in 29 off-street car parks located in Birmingham, UK.

In another comparative analysis, \citet{zhao2020comparative} reviewed various prediction methods, including linear regression, support vector machines (SVM), neural networks, and auto-regressive integrated moving average (ARIMA). Their evaluation, based on data from four car parks across three Chinese cities, also explored the impact of differentiating between weekdays and weekends. They found that SVM consistently provided stable and accurate predictions across nearly all car park types and sizes, while distinguishing weekdays from weekends improved results mainly for medium-sized and office-related parking lots.

Similarly, \citet{awan2020comparative} analysed sensor-collected occupancy data from Santander, a smart city in Spain, over a period of nine months. Their study concluded that simpler algorithms, such as random forests, systematically outperformed more complex deep learning models, such as neural networks.

All these approaches share the three limitations we aim to address in our work.

Several other works relax the assumptions regarding data requirements. However, these approaches still face important limitations in terms of interpretability and their ability to model unmet demand.

\citet{chen2014parking} consider different levels of data aggregation to predict parking space availability without relying on an explicit behavioural model. They analyse complex usage patterns, including multiple daily peaks, which are more intricate than those addressed in our work.

\citet{ji2015short} propose a wavelet neural network (WNN) model for short-term parking prediction, demonstrating higher accuracy and computational efficiency compared to traditional methods using real-world data from Newcastle, UK.

\citet{Chawathe2019UsingHD} introduce a data-driven predictive approach that leverages historical occupancy rates to estimate current and future availability. A key advantage of this method is that it requires minimal or no additional instrumentation of parking facilities, while still achieving sufficient accuracy for practical guidance.

\citet{fokker2022short} conduct long-term (6-months ahead) off-street occupancy forecasting for 57 parking facilities in Amsterdam. Their models incorporate external factors such as weather, local events, parking tariffs, and public transport changes. While this work incorporates rich contextual data, the models used offer limited interpretability.

\citet{VAKRINOU2025} address both short- and long-term occupancy forecasting across multiple off-street parking facilities in central Athens, using a multisource dataset that includes traffic, public transport operations, and weather conditions. Their approach employs LSTMs, Transformers, and XGBoost. Although XGBoost provides a degree of interpretability, the models are not grounded in an explicit behavioural framework, and cannot be used to reason about unmet demand.

\begin{table}[!t]
\caption{Overview of limitations in existing work on car park occupancy modelling. The table summarises whether each group of studies provides interpretable models, operates without requiring high-granularity data (e.g., individual arrival and departure times), and whether they can account for unmet demand. Our proposed framework addresses all three aspects.}
\label{tab:rel_work}
\begin{tabular}{@{}L{3.7cm} C{2.2cm} C{3.4cm} C{2.4cm}@{}}
\toprule
\textbf{Related work} & \textbf{Provides interpretable models} & \textbf{Does not require high-granularity data} & \textbf{Able to handle unmet demand} \\
\midrule
\citet{Zheng2015ParkingAP}, \citet{Camero2018EvolutionaryDL}, \citet{zhao2020comparative}, \citet{awan2020comparative}
& \xmark & \xmark & \xmark \\
\midrule
 \citet{chen2014parking}, \citet{ji2015short}, \citet{Chawathe2019UsingHD}, \citet{zhao2020comparative}, \citet{fokker2022short},  \citet{VAKRINOU2025}
& \xmark & \cmark & \xmark \\
\midrule
\citet{Tavafoghi2019AQA}, \citet{daniotti2023maximum}, \citet{schneble2025statistical}
& \cmark & \xmark & \xmark \\
\bottomrule
\end{tabular}
\end{table}

Some recent works introduce interpretable models, but they still rely on fine-grained data and are unable to reason about unmet demand.

\citet{Tavafoghi2019AQA} propose a queuing framework with non-homogeneous arrival rates for predicting occupancy in truck parking lots. Their method leverages individual arrival and departure times to estimate the service time distribution, assuming four non-overlapping heterogeneous populations of arrivals, each with a time-independent distribution. They also assume that no arriving vehicle finds the parking lot full, i.e., there are infinitely many parking spots.

\citet{schneble2025statistical} model on-street individual parking lot occupancy in the city centre of Melbourne as a two-state stochastic process using a semi-Markov framework. While the model is interpretable and statistically grounded, it requires detailed sensor-level data at the level of individual parking spots.

Similarly, \citet{daniotti2023maximum} analyses patterns in the number of parked car-sharing vehicles across specific areas in the Milan metropolitan region. They employ a Maximum Entropy modelling approach that enables the identification of rare or extreme events, such as weather anomalies or transit strikes. Despite its interpretability, the method also depends on high-resolution, disaggregated data.

Table~\ref{tab:rel_work} summarises the main limitations of the reviewed approaches along the three dimensions that motivate our work: model interpretability, ability to operate with aggregate data, and capacity to reason about unmet demand.

Our proposed framework is inspired by prediction techniques based on average activity profiles, such as those used in~\cite{kaltenbrunner2007description, szabo2010predicting}. Although these methods were originally developed for non-parking mobility data, they illustrate how temporal regularities in aggregate behaviour can be leveraged for forecasting. Similar ideas have been applied in the parking domain, where historical data has proven effective in predicting future occupancy levels even with limited data availability~\cite{Chawathe2019UsingHD}.

We end this section by reviewing conceptually related work which considers behavioural aspects of parking demand using questionnaire data. For example,~\citet{xue2019commuter} models commuter departure time choices under parking capacity constraints using bounded rationality principles. While their goal is not occupancy prediction per se, their approach highlights the role of limited parking availability in shaping temporal demand patterns and motivates the use of bell-shaped temporal distributions, a key component of our modelling framework.

Similarly, \citet{Zhao2017BehaviorDM} analyses parking behaviour in the context of Park-and-Ride facilities, using survey data to study how personal attributes, travel characteristics, and user intentions influence the decision to use such infrastructure. Their findings reveal that shortage of parking space is among the top three reasons users avoid Park-and-Ride, emphasising the importance of modelling not only observed demand but also unmet demand.

\section{Dataset Description}
\label{sec:dataset}
We use data provided by the metropolitan transport authority~\footnote{Autoritat del Transport Metropolit\`a~(ATM): \url{https://www.atm.cat/en/atm}.} of Barcelona (Catalonia) which has installed sensors in different park-and-ride parking locations with the aim to better understand their usage patterns and to develop a long-term strategy that allows to improve the interurban mobility from and to Barcelona. Each of the car parks is located in the vicinity of a regional train station, which allows people to park their cars and change directly there to the public transportation system, working as a bridge between both public and private transportation means \cite{Gen_strategy}.

\begin{figure}[t!]
\centering
\includegraphics[width=.9\textwidth]{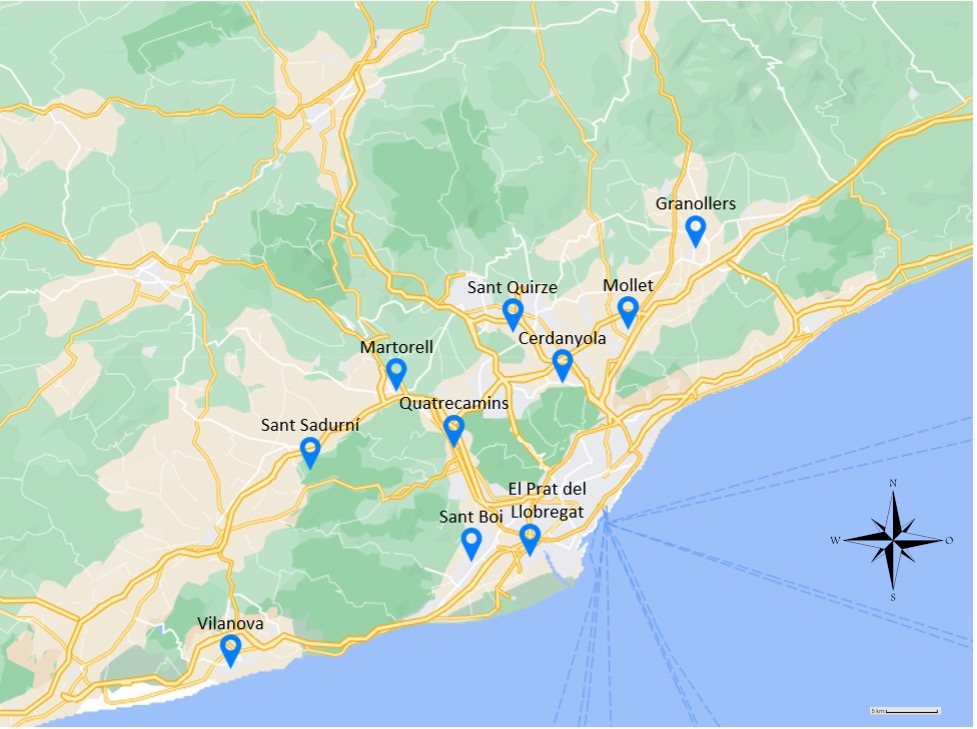}
\caption{Locations of the park-and-ride car parks analysed, all of them in the vicinity of Barcelona (Catalonia). Two of the locations, Martorell and San Quirze, have not been used due to errors in the data collection process. Scale indicates a 5km distance.}
\label{fig:location}
\end{figure}

Data from a total of 10 parking lots have been provided. Their geographical location is depicted in Figure~\ref{fig:location}. Data is continuously gathered from physical sensors placed in the entrances and exits of the car parks and stored in occupancy counters. The entrance of a vehicle increases the occupancy of the particular car park, and the departure of a vehicle decreases that counter. No data about the precise physical location where a vehicle is parked is recorded. This occupancy data is sampled in 30-minute intervals from every parking during a 3-month period from January 1st 2020, 0:00 to April 1st 2020, 0:00. Note that Catalonia entered a COVID-19-induced nationwide lockdown on March 15, 2020, and the car park usage decreased significantly. We will omit the data posterior to this date in the subsequent analysis. We have furthermore removed data from two car parks (Martorell and San Quirze) due to defective sensors and storage errors, which led to irregular readings.

\begin{figure}[t!]
\centering
\includegraphics[width=.8\textwidth]{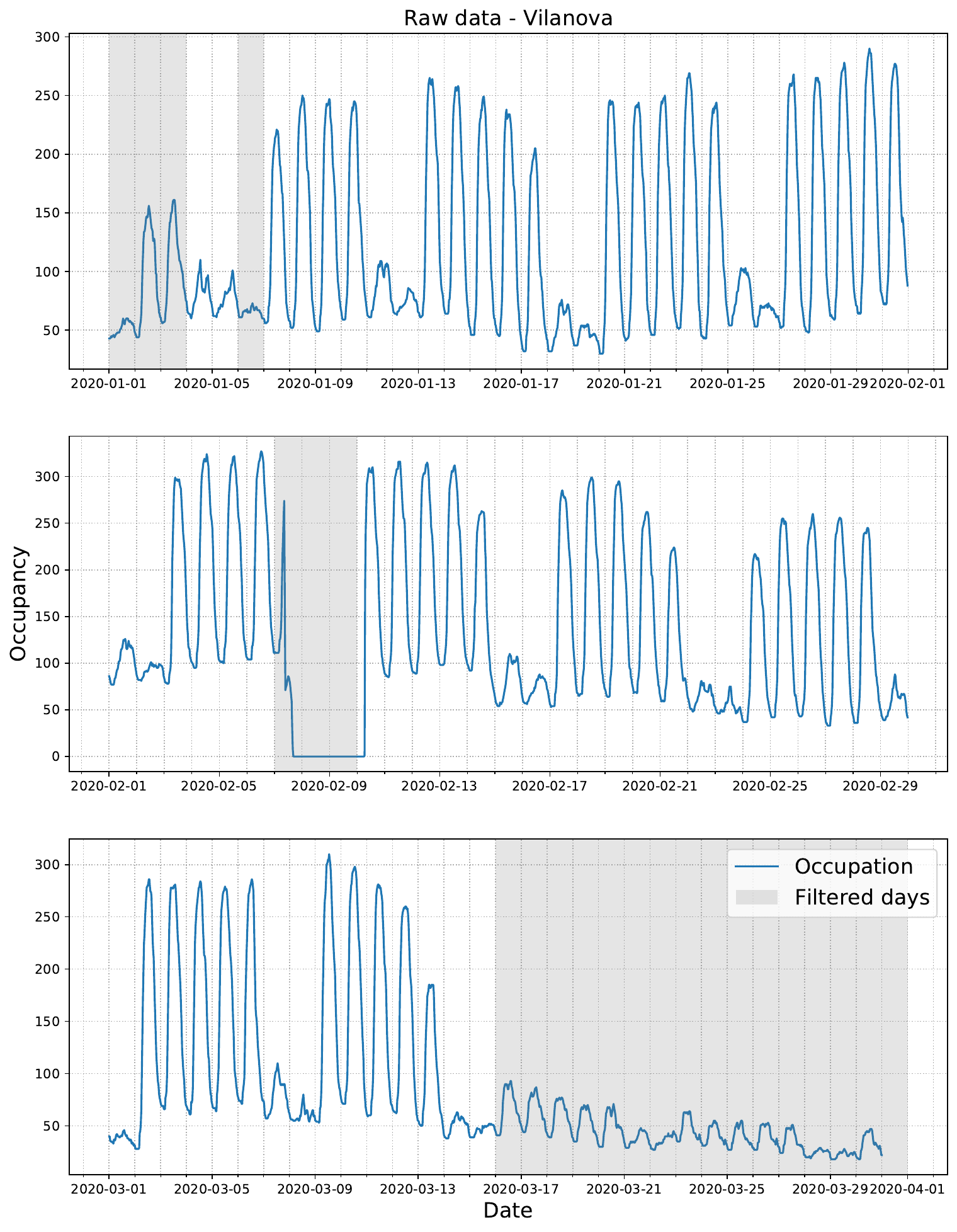}
\caption{Example of input data from the Vilanova car park, days with grey background have been removed from the analysis due to sensor failure (7th to 9th of February), COVID-19 induced lockdown (days after March 15th) or reduced activity due to Christmas holidays (1st to 3rd and 6th of January)}
\label{fig:rawdata}
\end{figure}

Figure~\ref{fig:rawdata} gives an example of the raw data from one of the car parks (Vilanova).
The observed occupancy daily pattern is the aggregate result of arrivals and departures to the car park location.
In the subsequent analysis, we have manually removed days where the occupancy profiles show errors due to sensor or storage failures. An example of such days are the days from February 7 to February 9 in Figure~\ref{fig:rawdata} or days that are influenced by holidays or vacation periods (e.g days 1 to 3 and 6 of January in Figure~\ref{fig:rawdata}, both periods together with the also removed COVID-19 lockdown period are shown with a grey background.

\begin{figure}[ht]
\centering
\includegraphics[width=.99\textwidth]{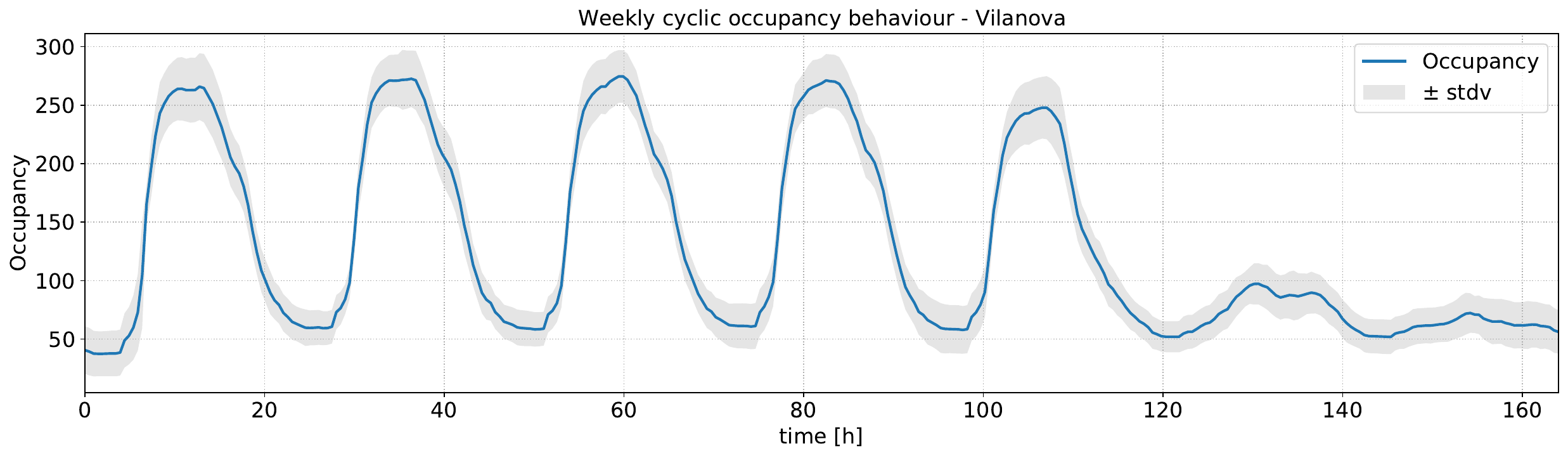}
\includegraphics[width=.99\textwidth]{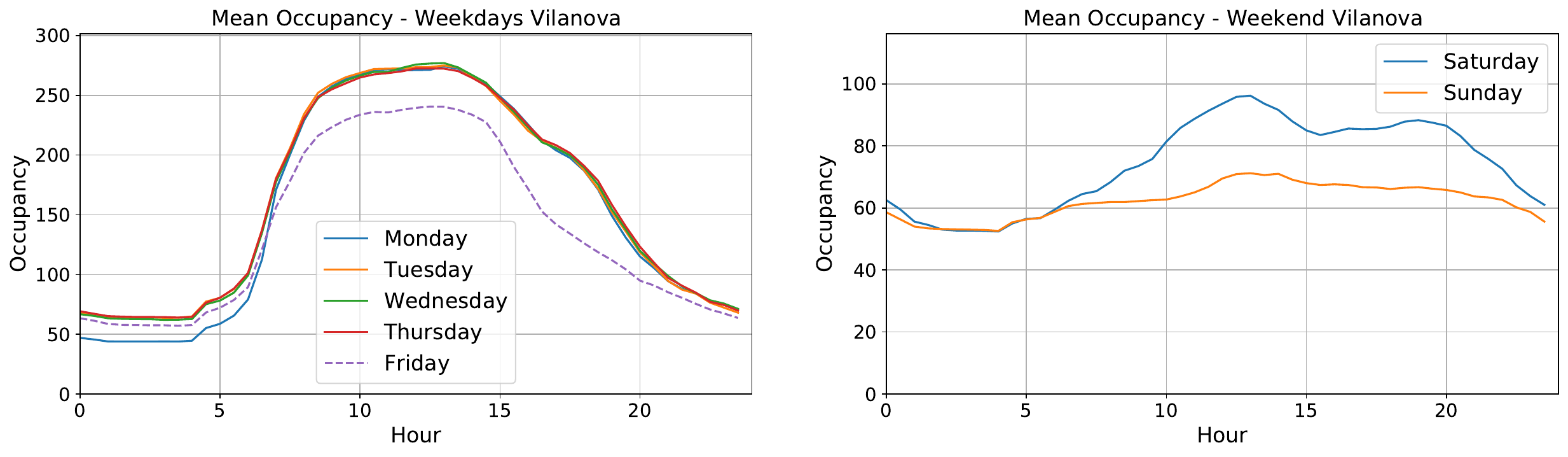}
\caption{Average weekly (top) and daily activity cycles (bottom) of the Vilanova car park. The daily average cycles are very similar from Monday to Thursday.}
\label{fig:cycle}
\end{figure}

This raw data already hints at the regularity in the parking behaviour. If we aggregate this data into weekly or daily activity cycles as is done in Figure
\ref{fig:cycle}, we can observe the regularity of the corresponding circadian patterns. The average behaviour from Monday to Thursday is nearly identical, and activity on Fridays shows a small decay.

In what follows, we aggregate the data in three groups: weekdays from Monday to Thursday, Fridays, and weekends (Saturdays and Sundays)



\begin{figure}[ht]
\centering
\includegraphics[width=.95\textwidth]{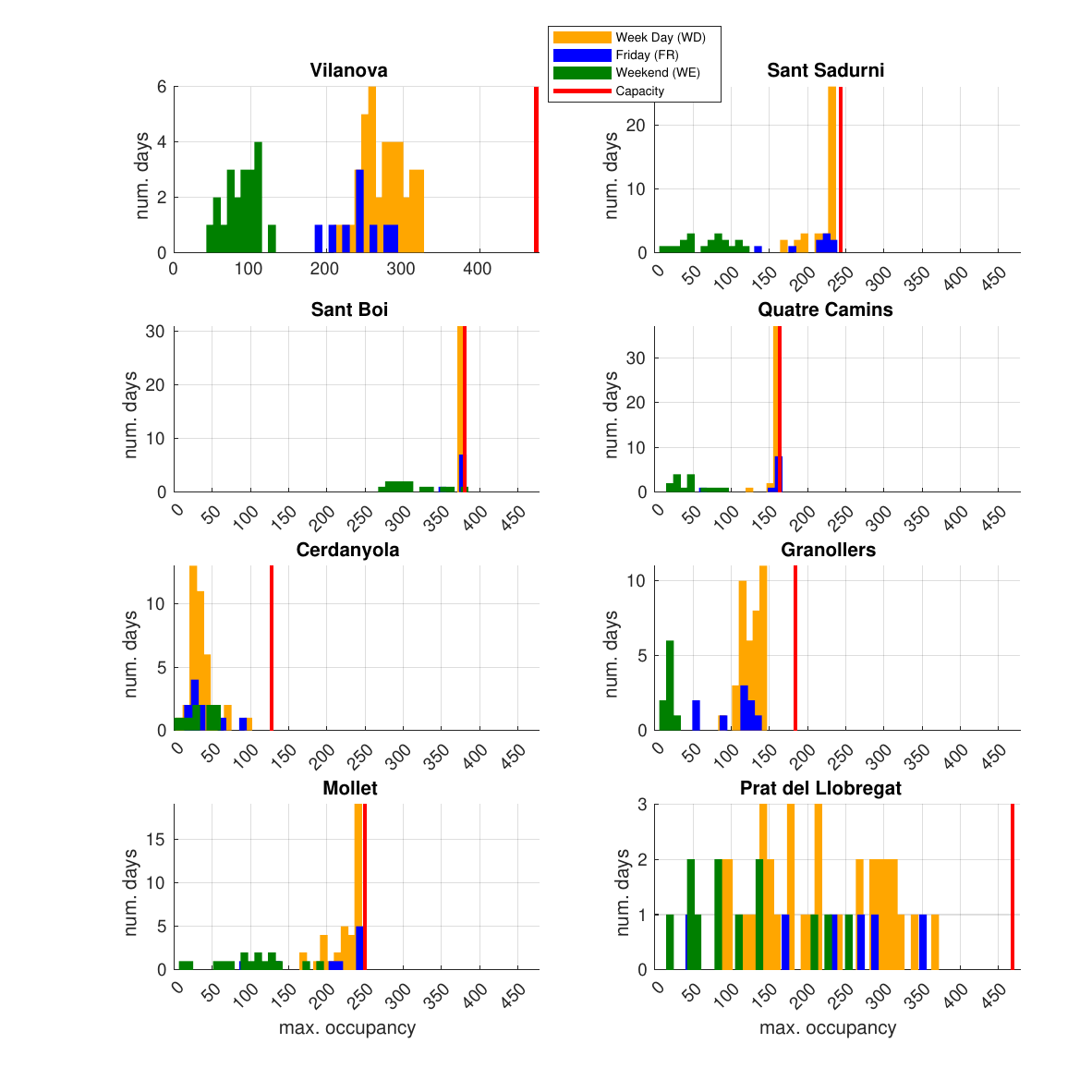}
\caption{Histogram of the observed maximal occupancies for the days between 01-01-2020 and 01-04-2020 aggregated by weekdays (orange), Fridays (blue) and weekends (green) for the eight car parks used in this study (bin sizes $= 9$ in all panels). Vertical red lines at the right of the panels indicate the maximal capacities of the car parks.
The maximum capacity was reached in SantSadurni (19 times), SantBoi (39 times) and in QuatreCamins (45 times).}
\label{fig:occupation}
\end{figure}

Finally, in Figure \ref{fig:occupation} we look at how close the different car parks get to their capacity limit (indicated by the red vertical line). For every day (excluding the days we have filtered out as mentioned above), we record the maximum occupancy of the car park and depict the corresponding distributions. Four of the car parks (Sant Sadurni, Sant Boi, Quatre Camins and Mollet) reach their maximal occupancy limit regularly during weekdays (orange bars) and also Fridays (blue bars). In the case of the example we used above, the Vilanova car park, this is not the case. We will use this distinction and develop a model extension for the stations where the capacity limit is reached.

\section{Modelling Car Park Occupancy}
\label{sec:model}
Given the regularity of the activity cycles we have seen in Figure
\ref{fig:cycle} we now present statistical models able to describe the daily occupancy patterns of an individual parking lot. 
The model considers the parking lot occupancy as a combination of two independent processes: arrivals and departures.
At any time, the occupancy is simply the aggregated number of arrivals minus the aggregated number of departures.

We introduce two types of models: a \emph{basic model} which assumes an unlimited amount of free parking spaces, and an \emph{extended model with occupancy limit}.
The former is suitable for parking lots which do not fill up completely, whereas the latter takes into account that a parking lot may reach its capacity limit, and cars arriving after this moment will have to find parking space elsewhere.

\subsection{Basic Model (Unlimited Parking Spaces)}
The proposed model accounts for two event types: car arrivals and departures. For simplicity, these events are assumed to be independent. Each process is modelled as a random variable following a (truncated) normal distribution.
The truncation is necessary to constrain the arrival and departure times to occur within the period corresponding to one day (which we normalise to the interval $[0,1]$).
The probability density functions for arrivals $\phi_a$ and departures $\phi_d$ are thus:
\begin{align}\label{eq:truncnorm}
\phi_a(t; \mu_a, \sigma_a) & = \frac{\phi(\frac{t-\mu_a}{\sigma_a})}{\sigma_a \cdot \left(\Phi\left(\frac{1-\mu_a}{\sigma_a}\right)-\Phi\left(\frac{-\mu_a}{\sigma_a}\right)\right)},\\
\phi_d(t; \mu_d, \sigma_d) & = \frac{\phi(\frac{t-\mu_d}{\sigma_d})}{\sigma_d \cdot \left(\Phi\left(\frac{1-\mu_d}{\sigma_d}\right)-\Phi\left(\frac{-\mu_d}{\sigma_d}\right)\right)},
\end{align}
where $\phi(\cdot)$ is the probability density function of the standard normal distribution and ${\displaystyle \Phi (\cdot )}$ its cumulative distribution function (CDF).
We denote the truncated CDFs as $\Phi_a(t; \mu_a, \sigma_a)$ for the arrival and  $\Phi_d(t; \mu_d, \sigma_d)$ for the departure process, which are determined by parameters $(\mu_a, \sigma_a)$ and $(\mu_d, \sigma_d)$ respectively.

At any time, the occupancy at the parking lot is given by the number of arrivals minus the number of departures.
We can generate a daily occupancy realization by drawing $M$ sample times from both processes $\{A_n\}_{n=1}^M \sim \phi_a$ and $\{D_n\}_{n=1}^M \sim \phi_d$
and subtracting the corresponding counting processes:
\begin{align}\label{eq:daily}
o(t) = \sum_{n\geq 1} \mathbbm{1}_{\{t\geq A_n\}} - \sum_{n\geq 1} \mathbbm{1}_{\{t\geq D_n\}}, \qquad t=[0,1],
\end{align}
where $\mathbbm{1}_{\{\cdot\}}$ is the indicator function that takes the value 1 when the condition is true and 0 otherwise. To know the occupancy of a car park at a given time $t$, we calculate the difference between all arrivals and departures that happened before.

A (normalized) continuous approximation to Eq.~\eqref{eq:daily} takes the difference between the CDFs of the arrivals $\Phi_a(\cdot)$ and 
departures $\Phi_d(\cdot)$: 
\begin{align}
f_\text{TN}(t; \theta) &= \Phi_a(t;\mu_a, \sigma_a) - \Phi_d(t;\mu_d, \sigma_d).
\end{align}


We name such a basic model the \textbf{Truncated Normal (TN) model}.
The TN model has parameters $\theta=(\mu_a,\mu_d,\sigma_a,\sigma_d)$, where $\mu_a$ and $\sigma_a$ denote the location time and the scale of the arrivals, respectively,
and $\mu_d$ and $\sigma_d$, which denote the average time and the scale of the departures, respectively.

\begin{figure}[!th]
\centering
\includegraphics[width=.9\textwidth]{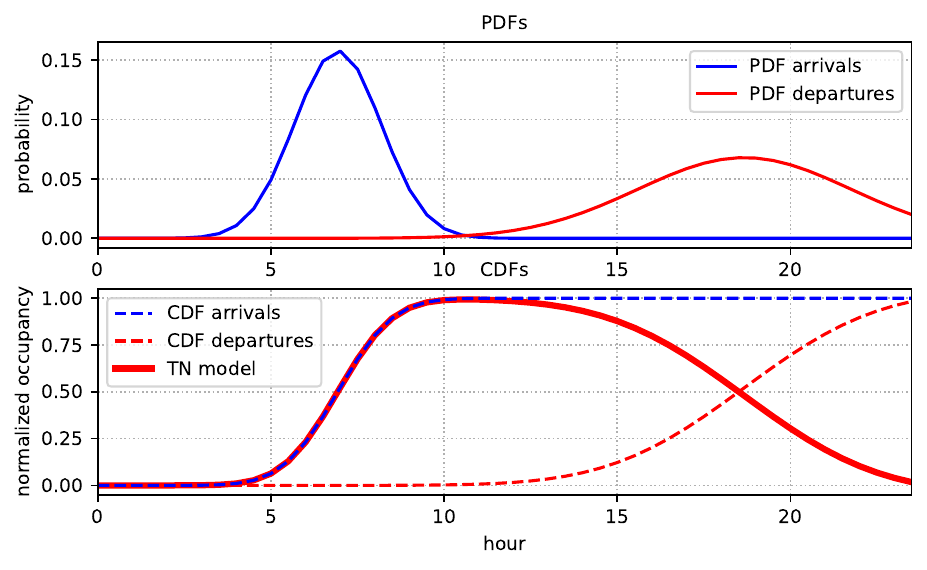}
\caption{Example of the TN model with data from the Vilanova car park. (top) probability density function (PDF) of car arrival (blue) and departure times (red), (bottom) corresponding CDFs (dashed lines) and car park occupancy curve (red thick solid line).   }
\label{fig:model}
\end{figure}

The top panel of Figure~\ref{fig:model} shows an example with a density function for arrivals (in blue) and departures (in red).
The bottom panel shows the corresponding CDFs (dashed lines) together with their difference (red solid line) as the aggregate result of cars parked in the parking lot.

The TN model provides a continuous temporal profile that can be easily adjusted if the available data consists of aggregated occupancies. Additionally, it provides a clear interpretability of the underlying mechanisms shaping the occupancy profile. Despite its simplicity, we will show that the TN model achieves predictive performance comparable to or superior to more complex approaches.


\begin{figure}[!b]
\centering
\includegraphics[width=.99\textwidth]{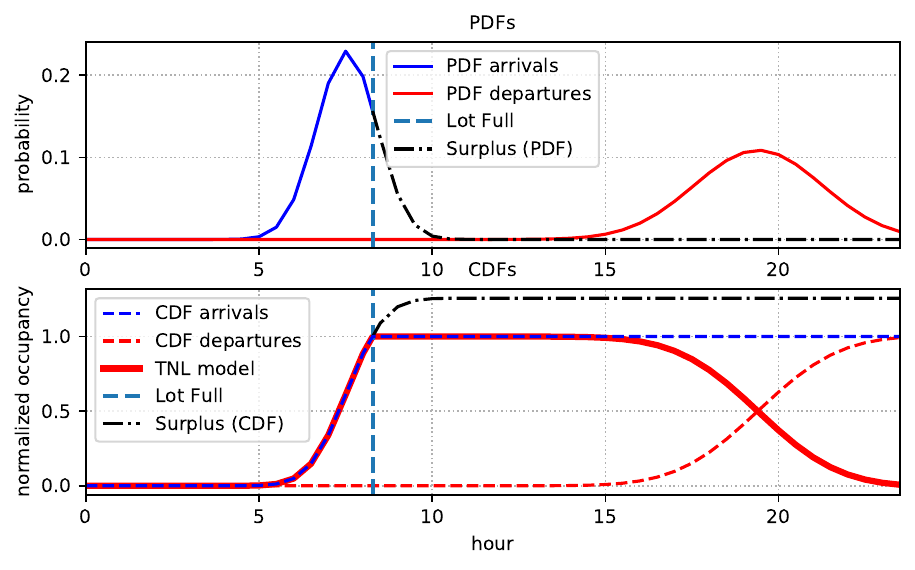}
\caption{Example of TNL model with data for the Quatre Camins car park. The car park fills up completely between 8 and 8:30 (cyan dashed vertical lines), which generates an excess of cars (dash-dotted black lines) which would arrive at the station but have to park elsewhere.}
\label{fig:model4c}
\end{figure}

\subsection{Model with Occupancy Limit}
We can extend the basic TN model to account for a capacity limit in the parking lot by incorporating a new parameter into the model.
Consider the total number of cars arriving in one day, and let $\tau$ be the fraction of those cars that fit in the parking lot.
We define the \textbf{Truncated Normal with Limit (TNL) model} as
\begin{align}
f_\text{TNL}(t; \theta) &= \bar{\Phi}_a(t;\mu_a,\sigma_a,\tau) - \Phi_d(t;\mu_d,\sigma_d),\notag\\
& \text{with }~\bar{\Phi}_a(t;\mu_a,\sigma_a,\tau)= \min\left(\frac{\Phi_a(t;\mu_a,\sigma_a)}{\tau},1\right)
\end{align}
where $\bar{\Phi}_a(\cdot)$ 
is a truncated and re-scaled version of $\Phi_a(\cdot)$ accounting for the fraction~$\tau$ only. In other words $\bar{\Phi}_a(\cdot)$ uses only the part of $\Phi_a(\cdot)$ which is smaller than $\tau$. 

The TNL model allows not only to characterise the moment $t_{L}$ when the capacity limit is reached (i.e. 
$\Phi_a(t_L;\mu_a,\sigma_a)=\tau$), 
but also to have an estimate of the number of cars which did not find a free parking space.
This is very relevant in scenarios where informed decision-making is required to determine policies to resize crowded parking lots.


Figure~\ref{fig:model4c} illustrates the TNL model for a day when the parking lot capacity limit was reached between 8 and 8:30 (vertical dashed line in cyan).
The top panel shows an example of a density function for arrivals (in blue) and departures (in red).
The arrival distribution is truncated when the parking fills up.
The dashed-dotted black line of the arrival distribution corresponds to the estimated density of cars arriving afterwards, which no longer fit in the parking lot.
The bottom panel shows the corresponding modified CDFs $\bar{\Phi}_a$ (dashed line in blue) and $\Phi_d$ (dashed line in red) together with the difference $\bar{\Phi}_a-\Phi_d$ (red solid line),
as the aggregate result of cars parked in the lot. Again, the proportion of surplus cars is indicated by the dashed-dotted line.

We will 
allow a different value for $\tau$ for every day $d$ of training data while using the same values for other parameters of the two distributions.
This way the parameters of the arrival distribution are independent of the actual moment a parking lot reaches its capacity limit and can be used,
not only to predict the moment when the capacity limit will be reached, 
but also to have an estimate of the number of cars which did not find a free parking space, as we will show later in Section~\ref{sec:nowcasting}. 

\subsection{Optimising Model Parameters from Observed Data}
In this section, we describe the procedure for estimating the model parameters using observed car park occupancy data.  

\subsubsection{Fitting the TN Model}  
For the basic TN model, the objective is to estimate the parameters  
$\boldsymbol{\theta} = (\mu_a, \mu_d, \sigma_a, \sigma_d)$ from a dataset consisting of $N$ daily occupancy profiles,  
$\mathcal{D}~=~\{o_{1:T}\}_{n=1}^N$, recorded at $T$ discrete timestamps\footnote{In our case, $T=48$ since the dataset has a 30-minute time resolution.}.  
To ensure consistency, all occupancy profiles are normalised such that $\sum_{i=1}^T o_i = 1$. 

If occupancy profiles exhibit stability within a group, meaning there is minimal variation across weekdays or weekends,  we can reasonably assume that occupancy at time $t$ is independently and identically distributed. For simplicity, we model occupancies as Gaussian-distributed with mean $f_{\boldsymbol{\theta}}(t)$ and variance $\beta^2$.  
The likelihood of the parameters $\boldsymbol{\theta}$ is given by  
\begin{align}
p(o_t^{(n)} | t, \boldsymbol{\theta}, \beta) & = \frac{1}{\beta\sqrt{2\pi}}\exp\left(-\frac{1}{2\beta^2}
\left(o_t^{(n)}-f_{\boldsymbol{\theta}}(t)\right)^2\right) \\
p(\mathcal{D} | \boldsymbol{\theta}, \beta) & = \prod_{n=1}^N \prod_{t=1}^T  p(o_t^{(n)} | t, \boldsymbol{\theta}, \beta).
\end{align}
The optimisation can be carried out by minimising the sum of squares loss
\begin{align}
\mathcal{L}^* = \min \sum_{n=1}^N \sum_{t=1}^T \left(o_t^{(n)}-f_{\boldsymbol{\theta}}(t)\right)^2,
\end{align}
and the variance is given by $\beta^2 = \mathcal{L}^* / NT$.

\subsubsection{Fitting the TNL Model}
For the TNL model with occupancy limit, the set of fitted parameters $\boldsymbol{\theta}$ also includes a set of threshold parameters $\{\tau_i\}_{i=1}^N$. 
Thus, we allow $\tau$ to take a different value for each training day $i$ while keeping the other parameters fixed. This approach ensures that the parameters of the arrival distribution remain independent of the exact moment a parking lot reaches capacity.  
As a result, the model can not only predict when the capacity limit will be reached but also estimate the number of cars that were unable to find a free parking space.  
Note, however, that this approach requires a different normalisation of the data.  
In this case, all occupancy profiles are scaled such that $\max(o_{1:T}) = 1$.  
This normalisation is essential for identifying the moments when the parking lots reach full capacity.  


\subsubsection{Test-Training Split}  
To train the model, we split the available data into training and test sets after the data-cleaning process.  
The test set is fixed at three weeks, while the remaining data is used for training.  
For most stations, this results in an approximate $30/70$ test-training split.


\subsection{Alternative Approaches}  
In this section, we briefly describe two alternative methods against which we compare our proposed modelling approaches.  
\subsubsection{Average Activity Profile}  
A straightforward method for predicting patterns influenced by circadian rhythms is the use of average activity profiles, as employed in~\cite{szabo2010predicting}.  
We adapt this approach by computing the average over normalised, aggregated data, distinguishing between three groups: weekdays (Monday to Thursday), Fridays, and weekends (Saturday and Sunday).  
\subsubsection{Linear Regression}  
As another baseline, we use a linear regression model to predict car park occupancy based on prior activity.  
Formally, given a dataset of $N$ normalized daily occupancy profiles,  
$\mathcal{D} = \{o_{1:T}\}_{n=1}^N$, we aim to predict $o_y$ using past observations $o_{1:x}$, where~$x < y$.  
Since occupancy profiles represent cumulative arrivals and departures, we apply the \texttt{diff} function to obtain changes in occupancy at each 30-minute interval.  
Thus, we define the transformed input as $\hat{o}_{1:x} = \{0, \mathtt{diff}(o_{1:x})\}$. 

The regression model optimises the following objective:  
\begin{align}
\mathcal{L}^*_{\text{lr}} = \min \sum_{n=1}^N \left(\sum_{t=1}^x \beta_t \hat{o}_t^{(n)} + \beta_0 - o_y^{(n)}\right)^2.
\label{eq:FitReg}
\end{align}

\section{Results and Discussion}  
\label{sec:results}
In this section, we first evaluate how well the model fits the data, followed by an analysis of its effectiveness in predicting or nowcasting parking lot occupancy at a given moment.  

\subsection{Model Fit Quality}  
We begin by assessing how well the model represents both the training and test data.
For this, we use the mean relative proportional error, computed relative to the maximum occupancy of a parking lot. For clarity, we group the test data into three categories based on observed activity patterns (as illustrated in Figure~\ref{fig:cycle}): weekdays (Monday to Thursday), Fridays, and weekends (Saturday and Sunday). The goodness of fit of the models in terms of the average loss per training day is given in Tables \ref{tab:lossTN} and \ref{tab:lossTNL} in Appendix \ref{sec:param}.

\subsubsection{Example Fits on Training Data} 
\label{sec:example_fits}

\begin{figure}[ht]
\centering
\includegraphics[width=.31\textwidth]{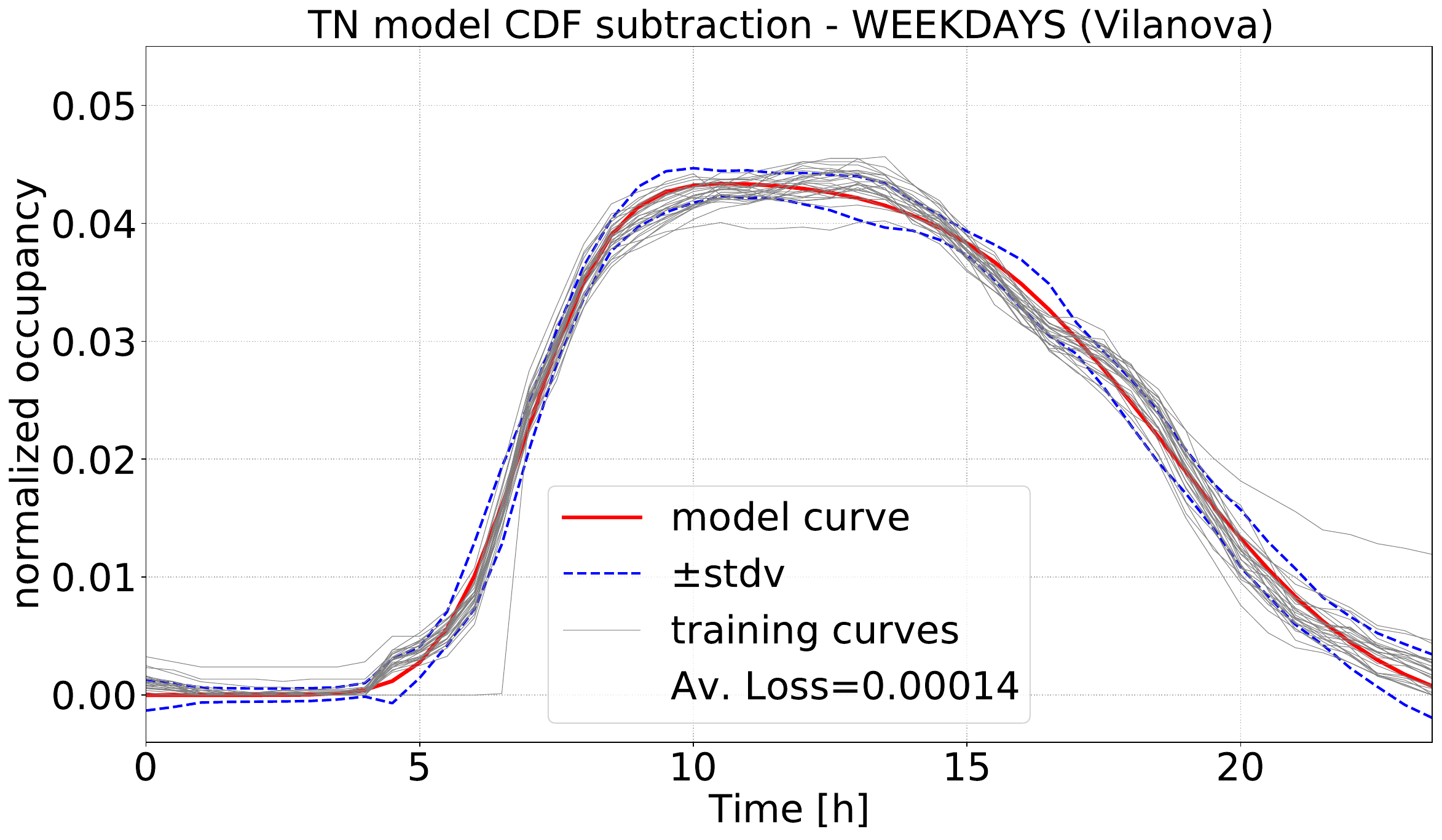}
\includegraphics[width=.31\textwidth]{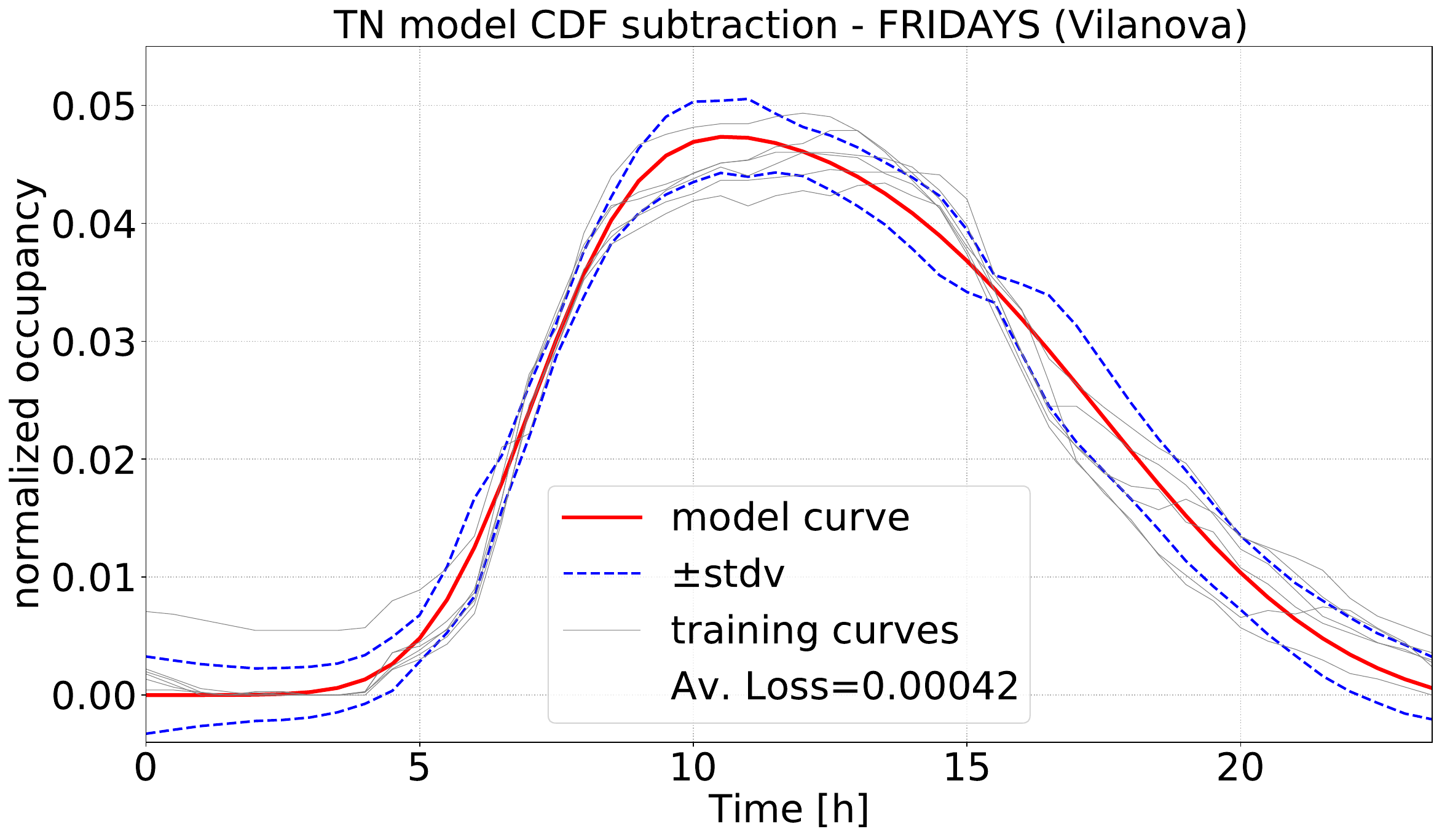}
\includegraphics[width=.31\textwidth]{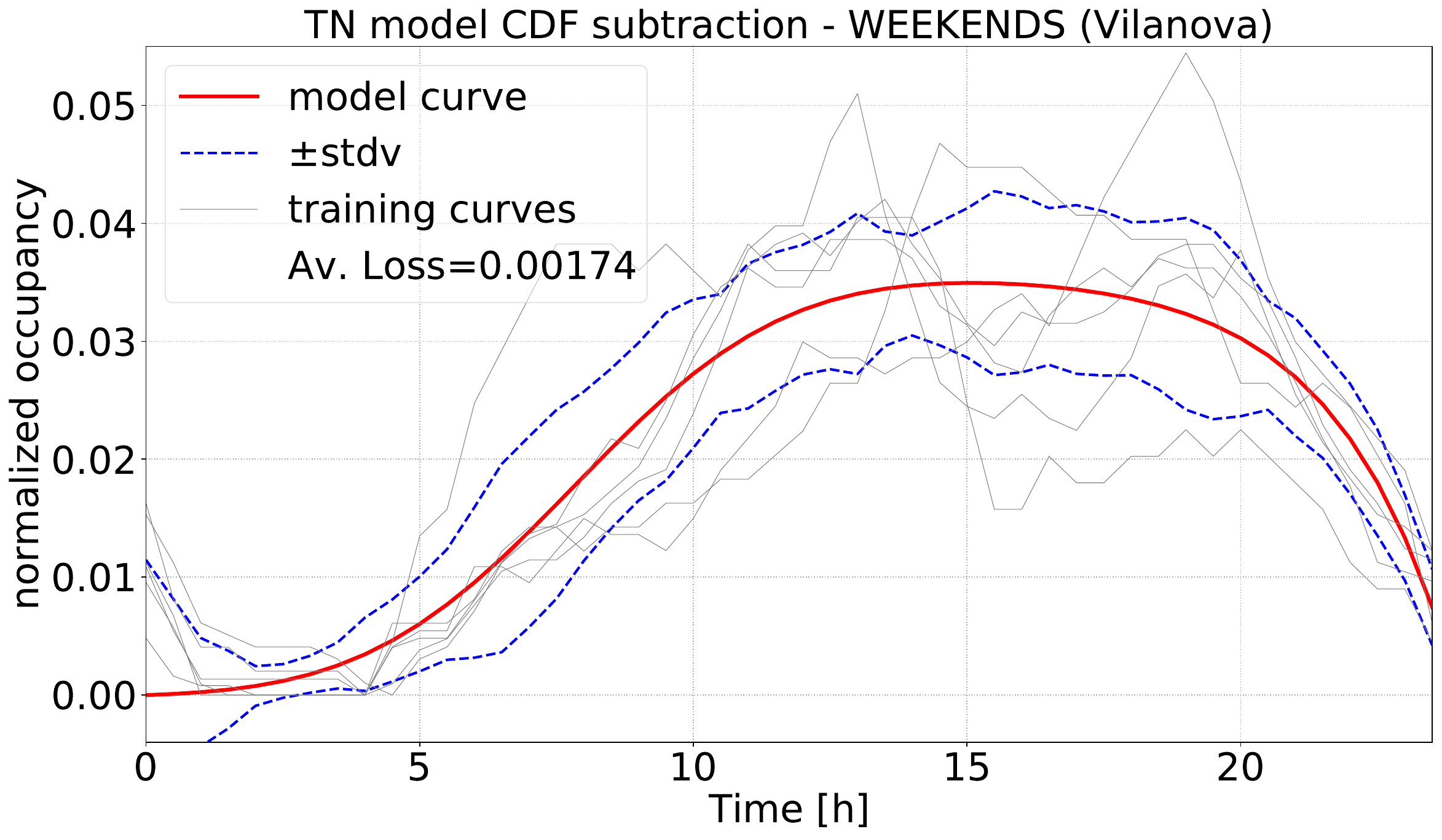}
\caption{Example of the TN model fitted to training data for the Vilanova car park. The  parameters of the arrival and departure TN distributions are $\mu_{a}=~$06:56h, $\mu_{d}=~$18:40h for weekdays, $\mu_{a}=~$07:02h, $\mu_{d}=~$17:27h for Fridays and $\mu_{a}=~$07:50h for weekends. Note that for the TN model the data is normalised by the area under the curve. Av. Loss indicates average sum of squared loss per day.}
\label{fig:trainingExmaple}
\end{figure}

\begin{figure}[ht]
\centering
\includegraphics[width=.31\textwidth]{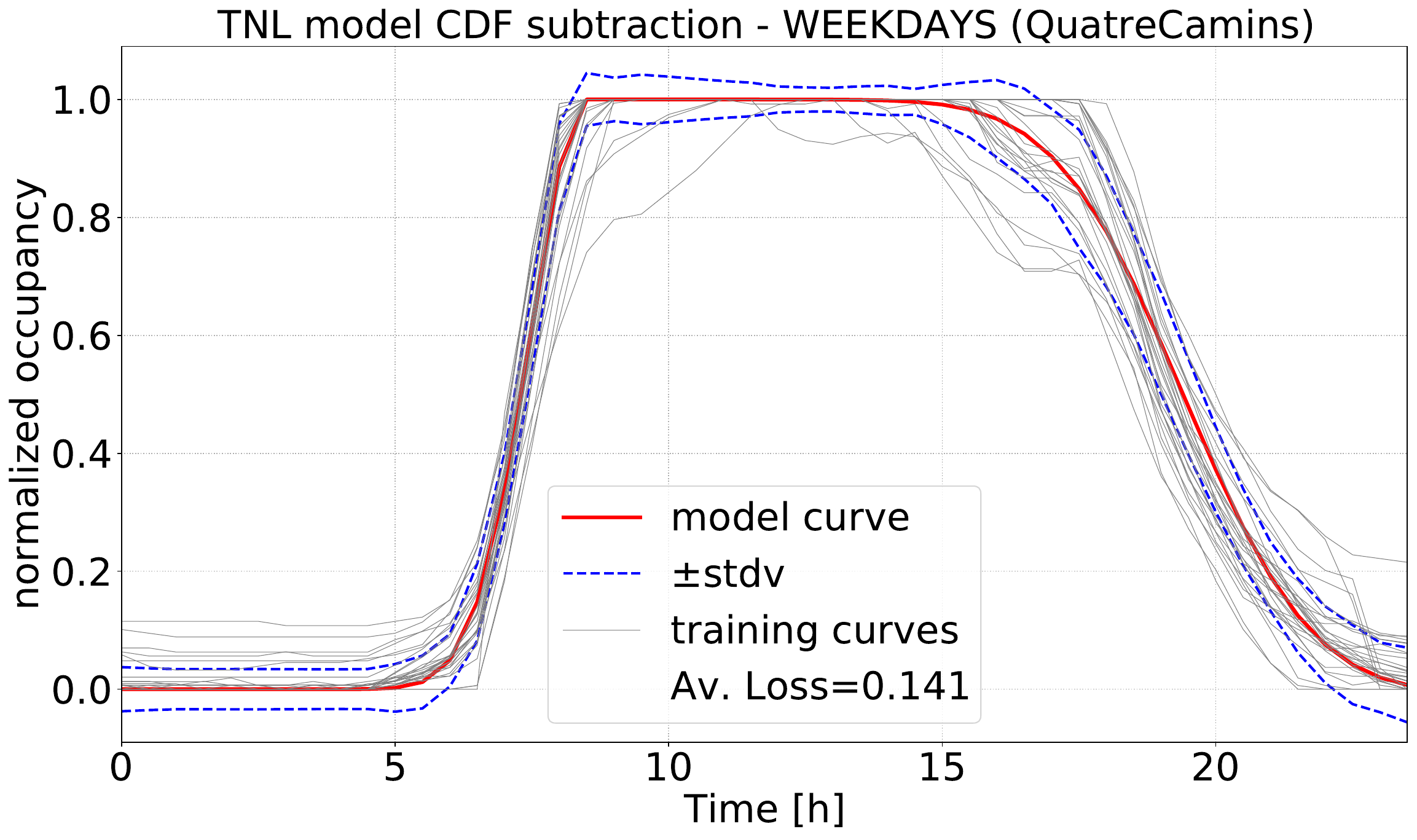}
\includegraphics[width=.31\textwidth]{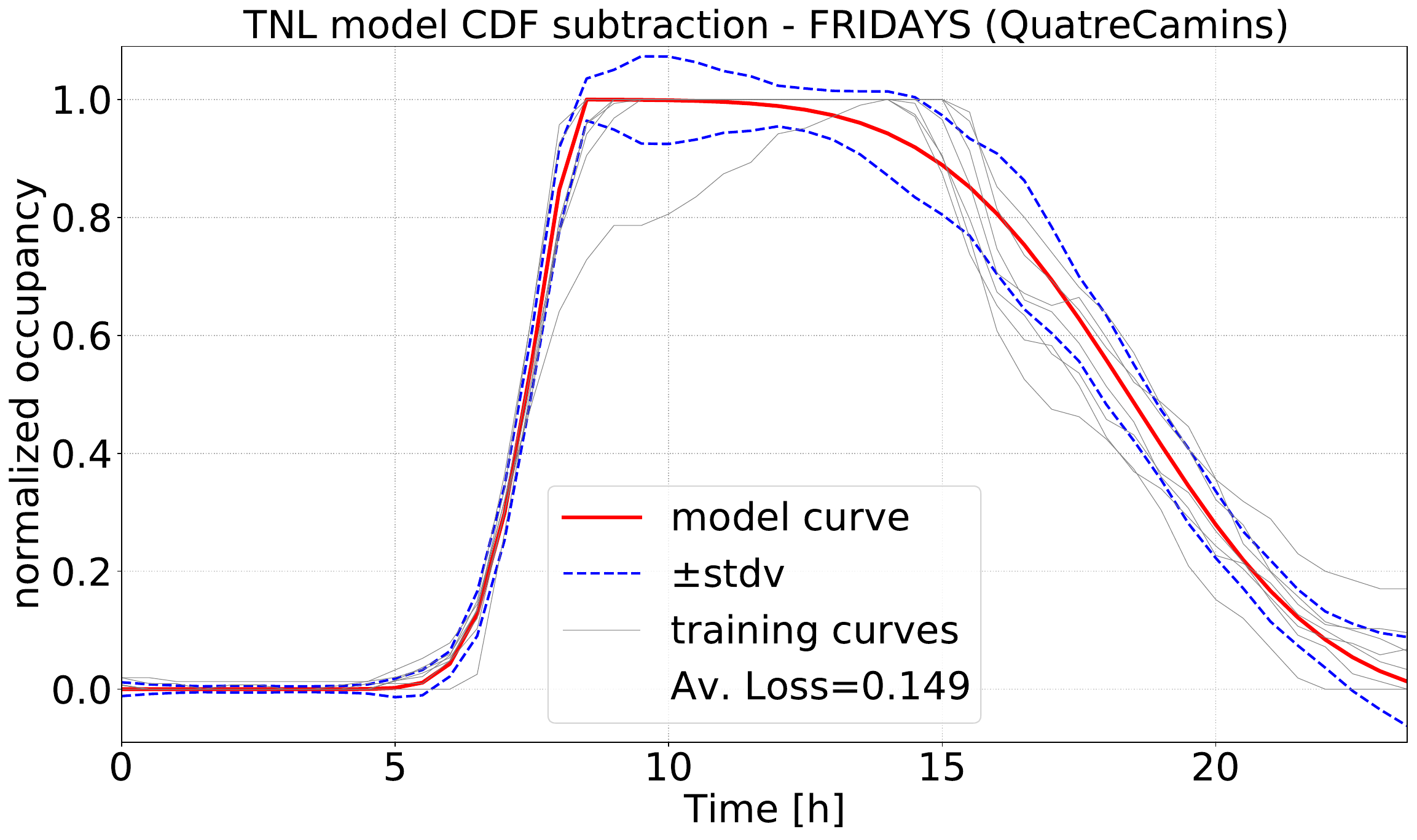}
\includegraphics[width=.31\textwidth]{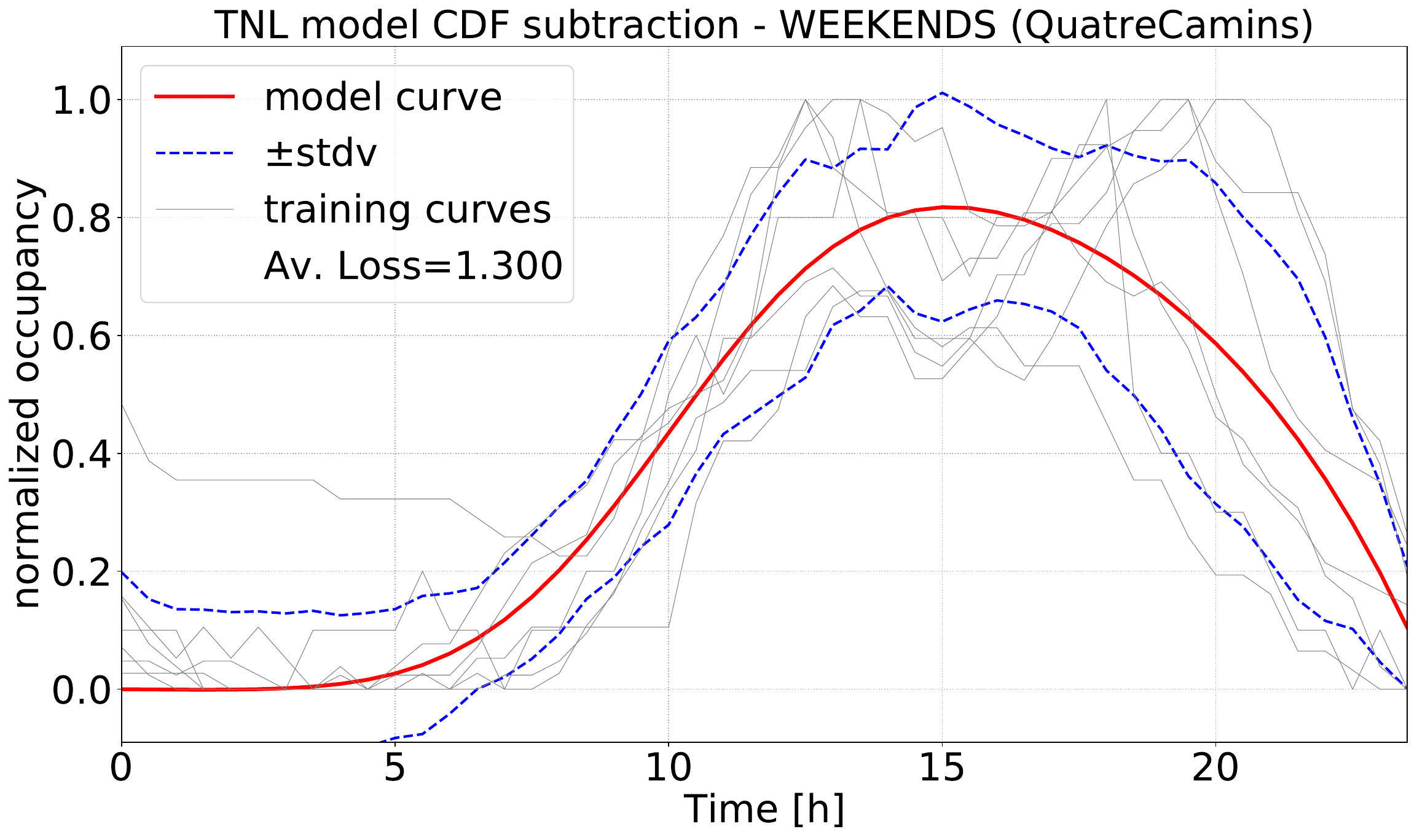}
\caption{Example of the TNL model fitted to training data for the Quatre Camins car park. The corresponding parameters of the arrival and departure distributions are $\mu_{a}=~$07:32h, $\mu_{d}=~$19:25h for weekdays, $\mu_{a}=~$07:43h, $\mu_{d}=~$18:30h for Fridays and  $\mu_{a}=~$10:10h and for weekends. Dashed lines indicate stdv of the model curve. The car park usually reaches its capacity limit from Monday to Friday between 8 and 8:30.  Note that for the TNL model the data is normalised by the max value.}
\label{fig:trainingExmaple4C}
\end{figure}

We first illustrate model fits on the training data using two example car parks.  
Figure~\ref{fig:trainingExmaple} shows the extracted TN model curves for the Vilanova car park, while Figure~\ref{fig:trainingExmaple4C}  
shows the TNL model curves, incorporating an occupancy limit, for the Quatre Camins car park.  
The latter reaches maximum capacity on most weekdays and Fridays (since for the weekend profile, this does not happen, the basic TN model is used there).

The dark grey lines represent actual occupancy profiles from the training data.  
The best-fitting model curves are shown as solid red lines, while blue dashed lines indicate the standard deviation of the model fit.  
All curves are normalised: in the TN model, by the area under the curve, and in the TNL model, by the maximum value.  
The latter normalisation is necessary to estimate excess demand in a given car park.  
For Figure~\ref{fig:trainingExmaple4C}, the model curve is plotted using the average excess across all training curves.  We observe the best fits for weekday profiles, while Fridays, especially weekends, exhibit greater variability.  

\begin{figure}[t]
\centering
\includegraphics[width=.45\textwidth]{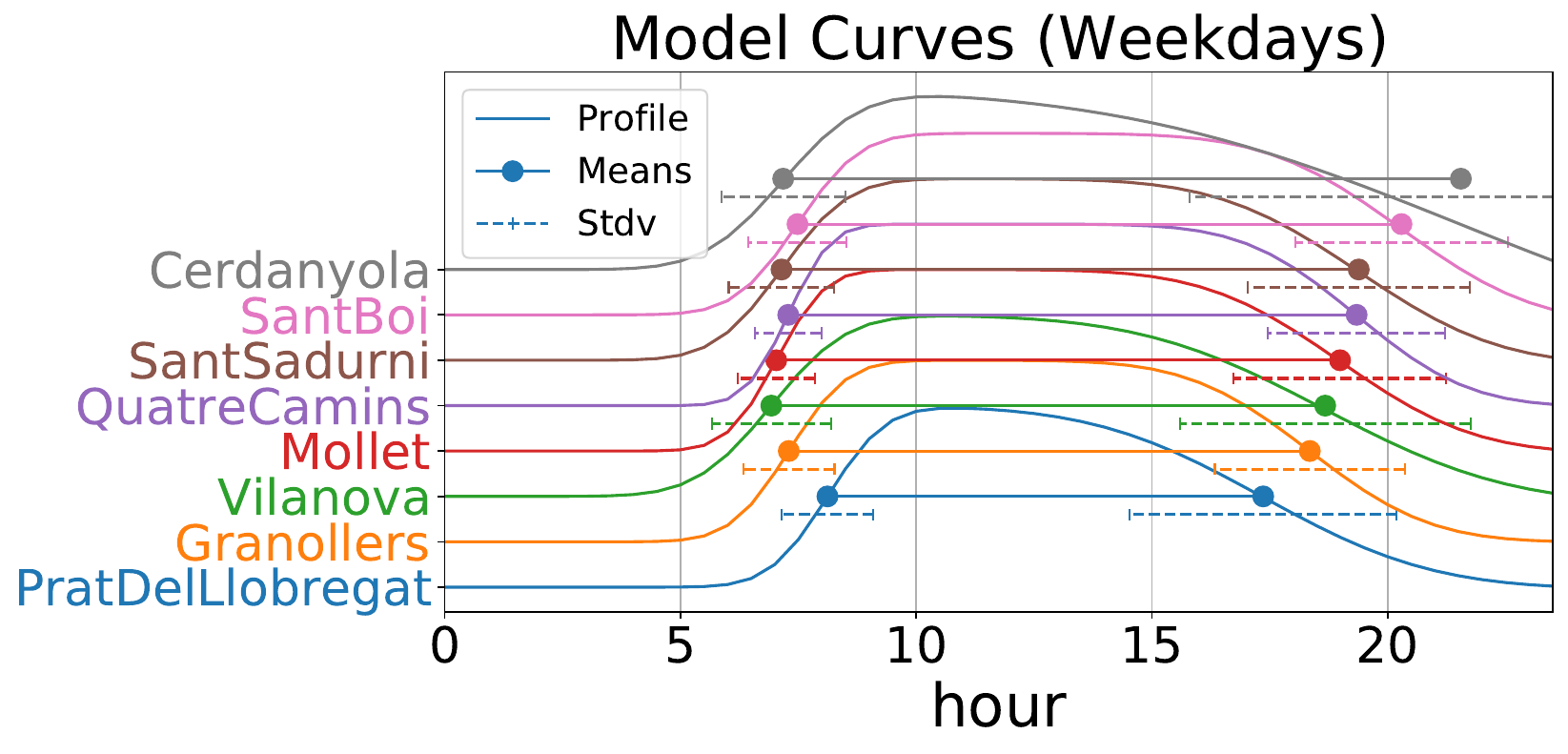}
\includegraphics[width=.45\textwidth]{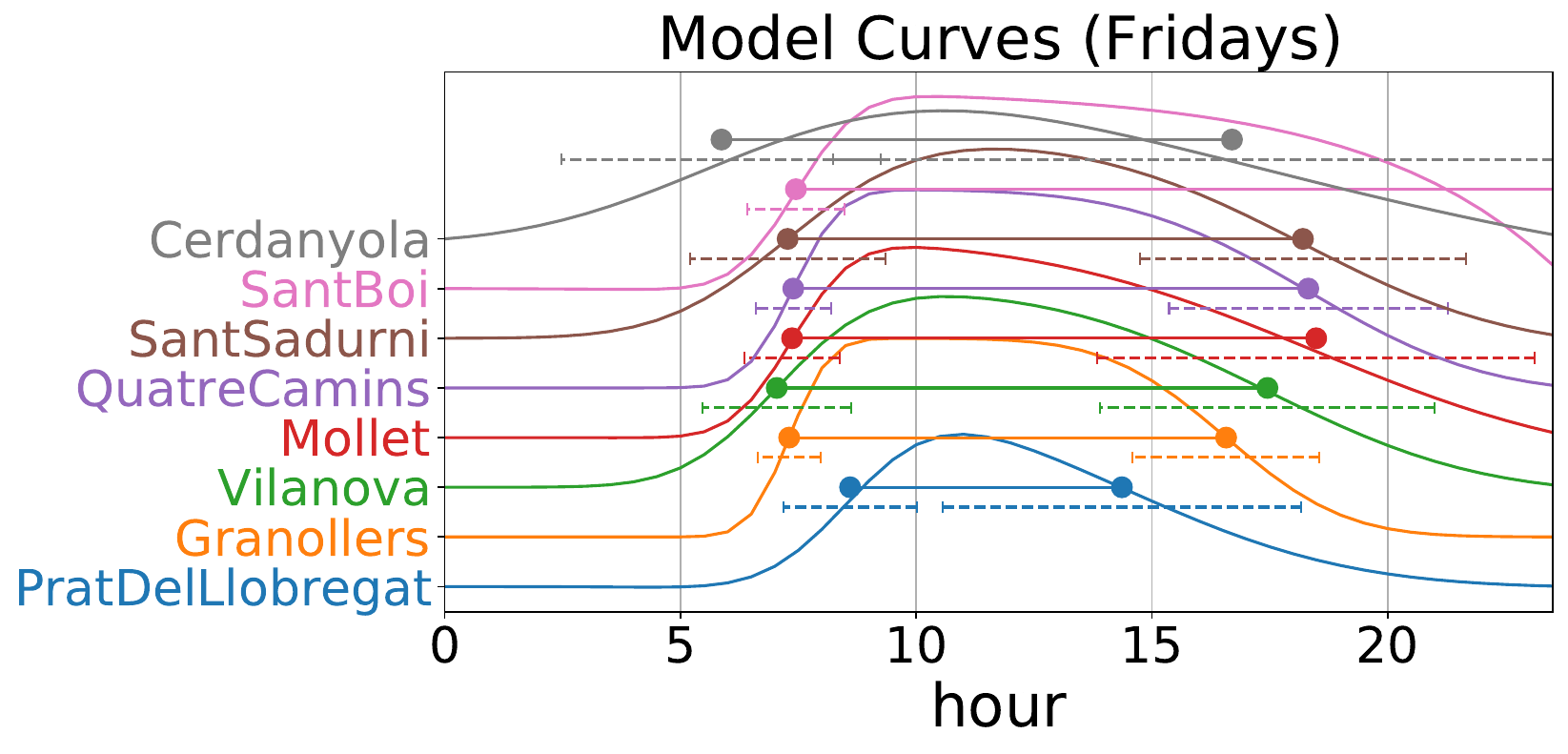}
\caption{Model curves, with parameters for the TN model, for weekends (left panel) and Fridays (right panel) for the different car parks. The dots indicate the mean $\mu$ and the dashed line the corresponding standard deviation interval $[\mu-\sigma,\mu+\sigma]$ of the underlying TN distributions for arrival (right) and departure (left).}
\label{fig:ModleCurvesParams}
\end{figure}

The fitted TN model curves and their corresponding parameters are shown in Figure~\ref{fig:ModleCurvesParams}.  
The figure presents model curves for weekdays (left panel) and Fridays (right panel) across all eight car parks, ordered by the difference between the location parameters $\mu$ of the arrival and departure distributions (represented as circle markers). The interval corresponding to $\pm$ one standard deviation $\sigma$ around $\mu$ is indicated by dashed lines and horizontal markers for both the arrival and departure distributions. The exact values of the parameters are given in Appendix~\ref{sec:param} in Table~\ref{tab:paramTN} for the TN model and in Table~\ref{tab:paramTNL} for the TNL model.


A smaller time difference between arrivals and departures is observed on Fridays, primarily due to earlier departure times.  
This pattern aligns with the common practice of shorter work schedules on Fridays, often without lunch breaks. The most pronounced example is the Prat del Llobregat car park, where the average departure time shifts from approximately 17:20 to 14:20—a three-hour advancement. We exclude the corresponding curves for weekends, 
which are characterised by more extreme parameter values and greater variability.  


\subsubsection{Evaluating Model Performance on Test Data}
\label{sec:evaluation}

We now analyse the model’s performance by evaluating errors in the test data. Examples of two individual stations are given in Appendix~\ref{sec:A1}.



\begin{figure}[!hb]
\centering
\includegraphics[width=.9\textwidth]{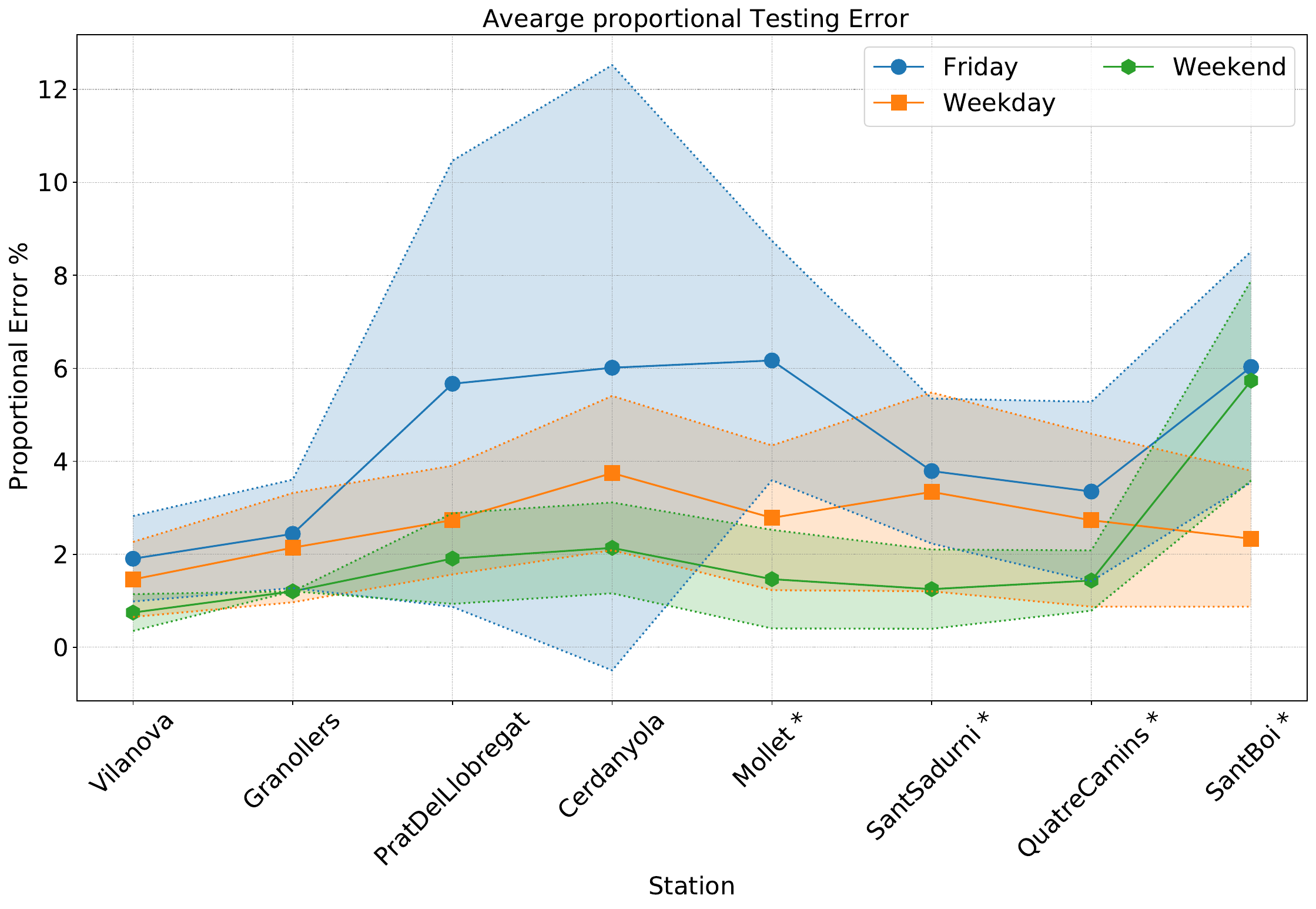}
\caption{Average proportional errors for different types of days of fitting the trained base model to test data for the 8 carparks with sufficient data. Shaded areas indicate the average standard deviation of the error measure. Stations with * use the model with occupancy limit (TNL) for weekdays and Fridays.}
\label{fig:testingErrorallStations}
\end{figure}

Taking the averages of these errors on the testing days and aggregating them into the three weekday groups, we obtain the errors depicted in Figure~\ref{fig:testingErrorallStations}. The four rightmost stations reach their occupancy limit during weekdays (including Fridays), and for these, we depict the error of the TNL model for the weekday and Friday profiles. The shaded areas represent the average daily standard deviation of the prediction errors.

We find that, apart from the four stations that show significant variability on Fridays (Prat Del Llobregat, Cerdanyola, Mollet, and Sant Boi, the latter also on weekends), the average error is always below 4\%. The high error on weekends for the Sant Boi car park is caused by party-goers visiting nearby nightlife attractions. This results in a high-occupancy phase during the early morning hours, which is not well captured by the current models. To address this issue, future work could extend the model and use cycles that last longer than 24 hours on weekends and end when the car parks reach their minimum occupancy.


In general, the models perform better on weekdays from Monday to Thursday, which exhibit more regular activity compared to Fridays. The error is nearly always lower for the weekend profile, but this is due to the lower activity on these days relative to the stations' maximum occupancy.


\subsection{Prediction}

Next, we analyse the predictive quality of our model. More precisely, at a given moment in time, we aim to predict the evolution of occupancy in the subsequent hours of the day. Again, we measure prediction quality using the relative absolute error with respect to the maximum occupancy of a car park and aggregate weekdays into three groups.
We also compare the results with average profiles, which represent the typical activity in a given car park for the corresponding types of days in the training data.

We make predictions by adjusting the model and the average profile to match the observed occupancy at the initial time. To achieve this, we simply shift and rescale the corresponding curves to best fit the known data. Formally, given an occupancy profile $o_{1:T}$ at a certain hour of the day $h < T$, we determine the parameters $\beta_0$ and $\beta_1$ that minimize  
\begin{align}
\mathcal{L}^* = \min \sum_{t=1}^h \left(o_t - \beta_0 - \beta_1 f(t)\right)^2,
\label{eq:FitPrediction}
\end{align}
where $f(t)$ represents either a model or the average activity profile, and $\beta_0$ corresponds to the number of cars parked at the station at the beginning of the day.

In the case of the TNL model with an occupancy limit, only the arrival distribution $\Phi_a$ is used for fitting. This allows us to calculate the excess number of cars, resulting in the following procedure:  
\begin{align}
\mathcal{L}^* = \min \sum_{t=1}^{\min(h,t_m)} \left(o_t - \beta_0 - \beta_1 \Phi_a(t;\mu_a,\sigma_a) \right)^2,
\end{align}
where $t_m$ is the time slot at which $o_t$ first reaches its peak value:
$$t_m= \min\{t \mid o_t = \max(o_{1:h})\}$$
In other words, we only use the phase up until the car park reaches its maximum capacity for fitting.

To determine the fraction $\tau_i$ of the arrival distribution that fits within the car park, we solve:
\begin{align}
\tau_i=\max\{\Phi_a(t;\mu_a,\sigma_a) \quad|\quad
\beta_0 + \beta_1 \Phi_a(t;\mu_a,\sigma_a,\tau_i) \leq \max(o_{1:T})\}.
\end{align}
This can be simplified to:
\begin{align}
\Phi_a(t;\mu_a,\sigma_a) \leq \frac{\max(o_{1:T}) - \beta_0}{\beta_1}
\end{align}
which leads to:
\begin{align}
\tau_i = \frac{\max(o_{1:T}) - \beta_0}{\beta_1}.
\end{align}
The departure distribution \( \Phi_d(t;\mu_d,\sigma_d) \) is then rescaled as follows:
\begin{align}
\bar{\Phi}_d(t;\mu_d,\sigma_d) = \Phi_d(t;\mu_d,\sigma_d) \cdot (\max(o_{1:T}) - \beta_0).
\end{align}
Essentially, this ensures that the departure distribution returns to a baseline of \( \beta_0 \) cars at the end of the day.

The excess, i.e., the number of cars that the arrival distribution predicts will arrive at the parking lot after it reaches full capacity, can be calculated as
\begin{align}
\mathrm{Excess}(T) = \beta_1 \cdot \max(\Phi_a(t;\mu_a,\sigma_a)) + \beta_0 - \max(o_{1:T}) = \beta_1 + \beta_0 - \max(o_{1:T}).
\end{align}

\begin{figure}[ht]
\centering
\includegraphics[width=.9\textwidth]{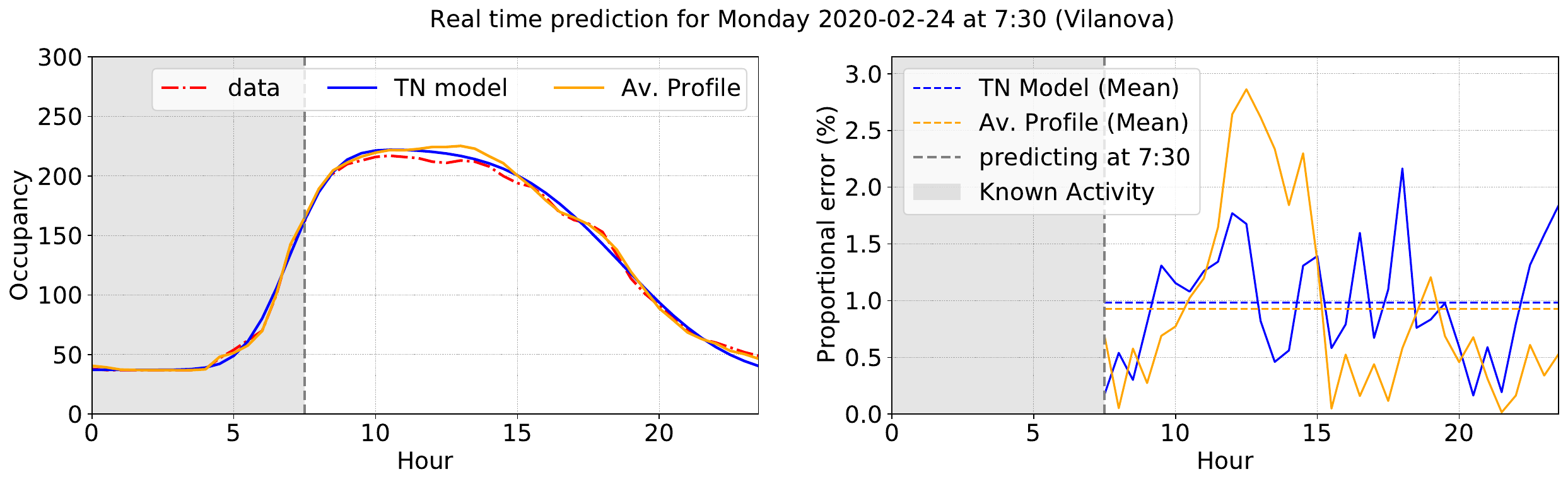} 
\includegraphics[width=.9\textwidth]{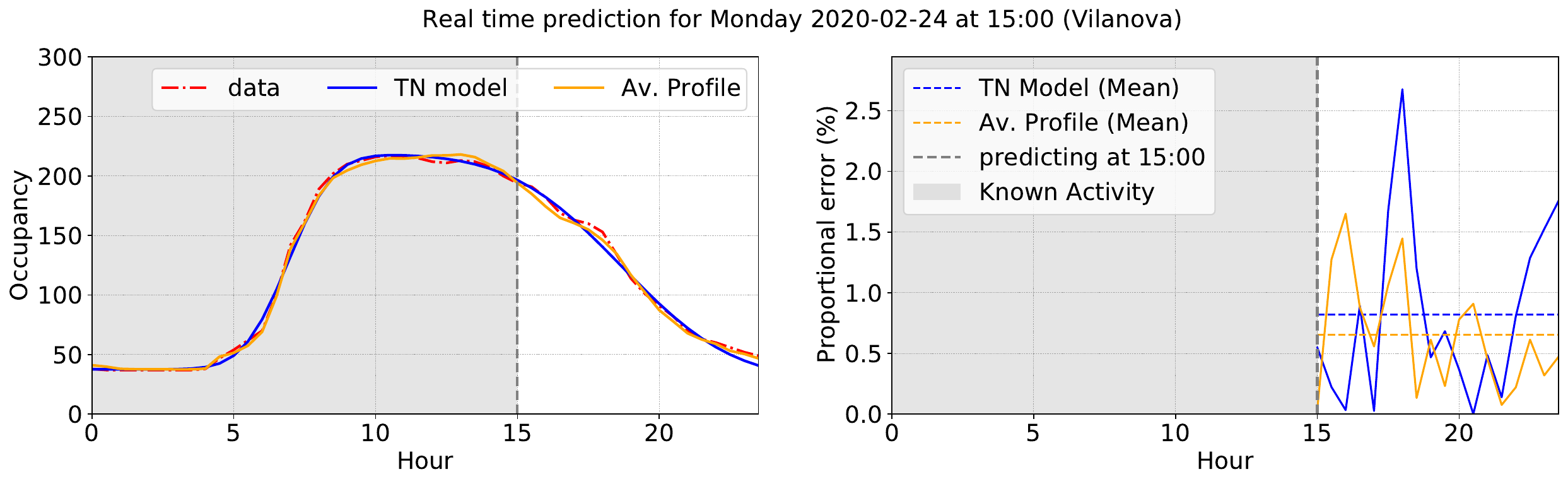}
\includegraphics[width=.9\textwidth]{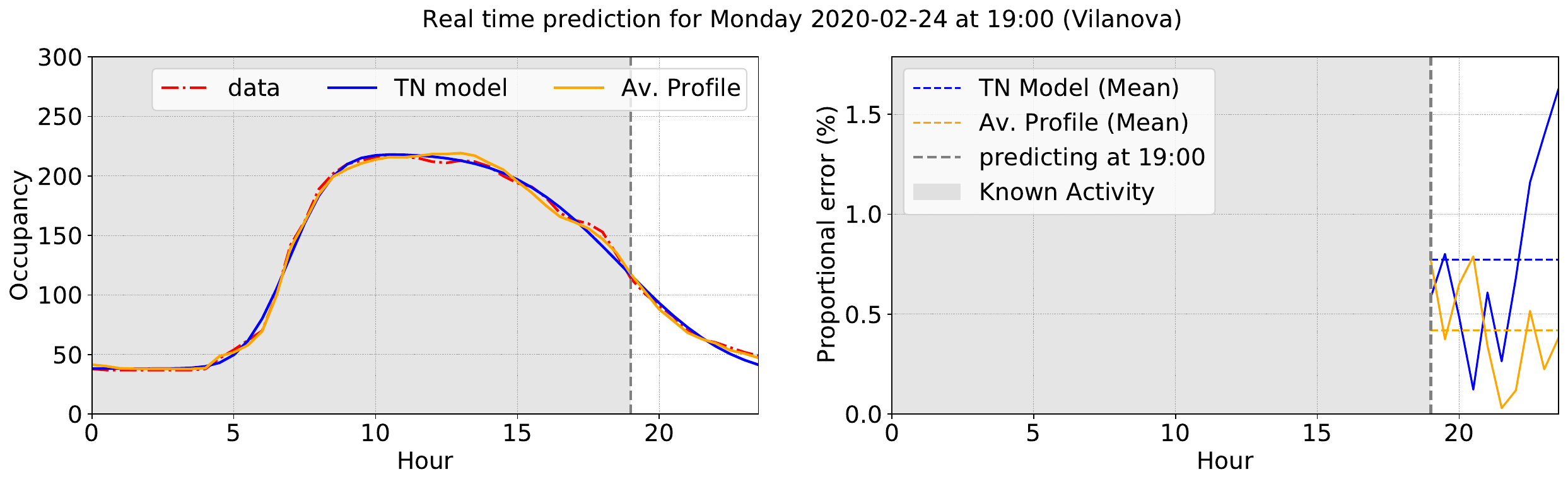}
\caption{Three examples of predictions in a car park (Vilanova) with the TN model. Left: The grey areas indicate the known activity used to predict the number of parked cars (red dashed-dotted lines) during the remaining hours of the day. Right: evolution of the relative prediction error in \%. Blue lines indicate performance of the TN model, while orange lines show the comparison with an average day-cycle profile. }
\label{fig:PredictionExampleThVilanova}
\end{figure}

\begin{figure}[ht]
\centering
\includegraphics[width=.9\textwidth]{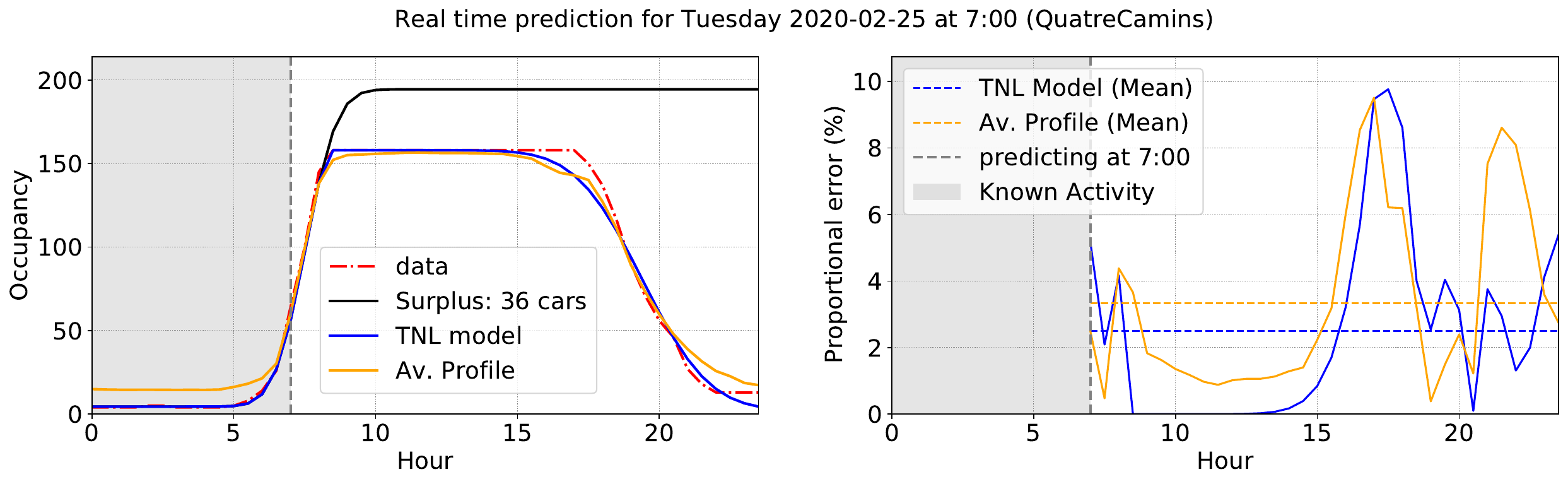}
\includegraphics[width=.9\textwidth]{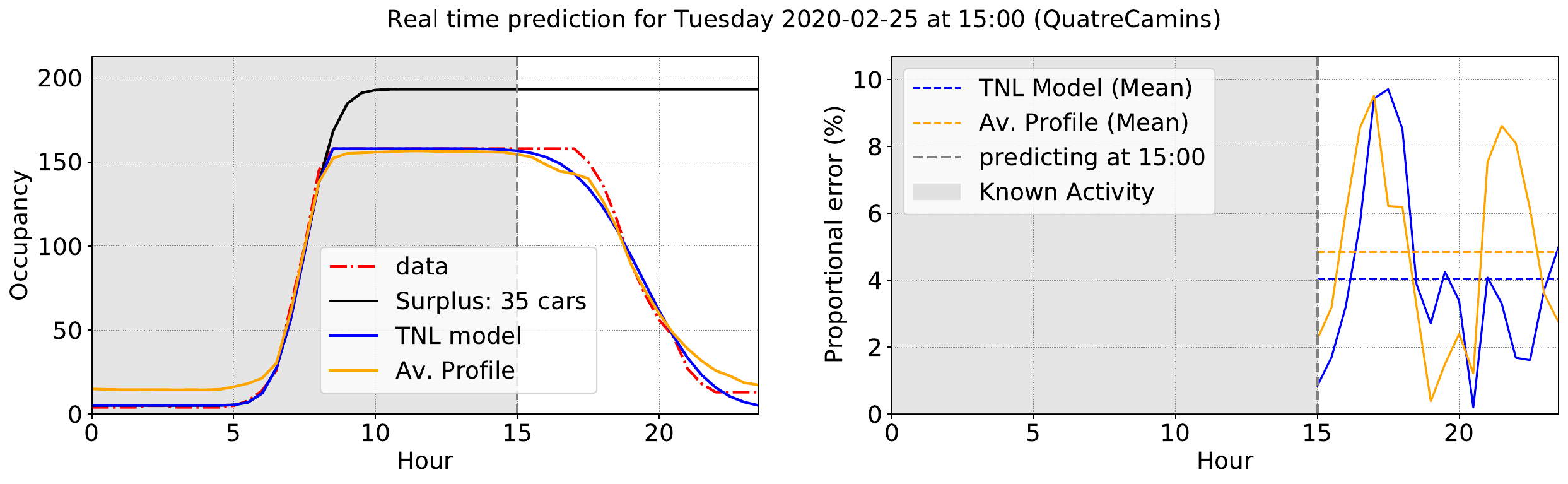}
\includegraphics[width=.9\textwidth]{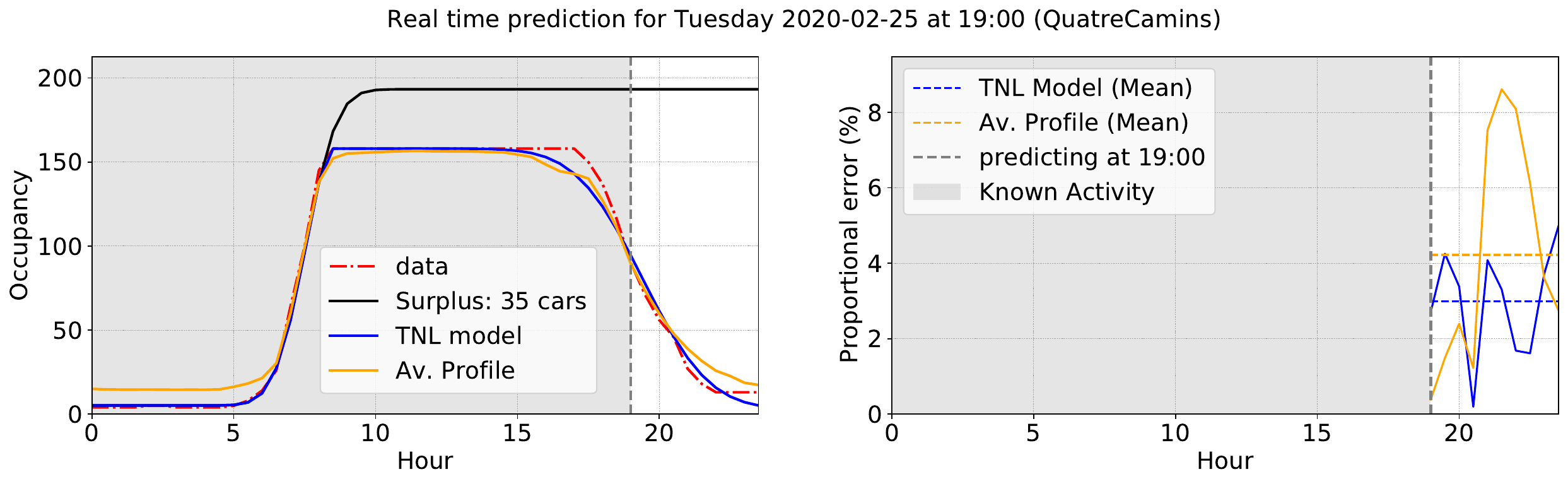}
\caption{Three examples of predictions in a car park which fills up (QuatreCamins) with the TNL model. Left: The grey areas indicate the known activity used to predict the number of parked cars (red dashed-dotted lines) during the remaining hours of the day. Right: evolution of the relative prediction error in \%. Blue lines indicate performance of the TNL model, while orange lines show the comparison with an average day-cycle profile. Black line gives an estimate of the surplus (how many cars will not or did not fit in the parking).}
\label{fig:PredictionExampleTh}
\end{figure}

Figures~\ref{fig:PredictionExampleThVilanova} and~\ref{fig:PredictionExampleTh} provide examples of predictions performed at 7:00, 15:00, and 19:00 for the Vilanova (TN model) and Quatre Camins (TNL model) car parks.  

The left panels display real data (dash-dotted lines) along with the predicted occupancy, while the right panels show the corresponding prediction errors for both the model curve (in blue) and the average profile (in orange). The average errors are depicted as dashed lines. The dark grey area represents the data used to generate the prediction.

Figure~\ref{fig:PredictionExampleTh} also includes an estimate of the surplus of cars that could not fit in the car park, indicated by the black continuous line. We observe that the performance of the TNL model is slightly better than that of the average profile for the Quatre Camins car park, and the surplus prediction remains stable across the three depicted time points.

\subsection{Nowcasting}
\label{sec:nowcasting}
To further evaluate the model’s performance, we apply it to a nowcasting task, i.e., we assess how well it predicts the number of cars in the car parks one hour ahead.

We quantify this by measuring the following prediction error, using the parameters obtained in Eq.\eqref{eq:FitPrediction}:
\begin{align}
\mathcal{E} = \sum_{t=h}^{h+w} \frac{|o_t - \beta_0 - \beta_1 f(t)|}{w \cdot \max(o_{1:T})},
\end{align}
where $w = 2$ (corresponding to a prediction window of 1 hour with a 30-minute data resolution).

\begin{figure}[ht]
\centering
\includegraphics[width=.45\textwidth]{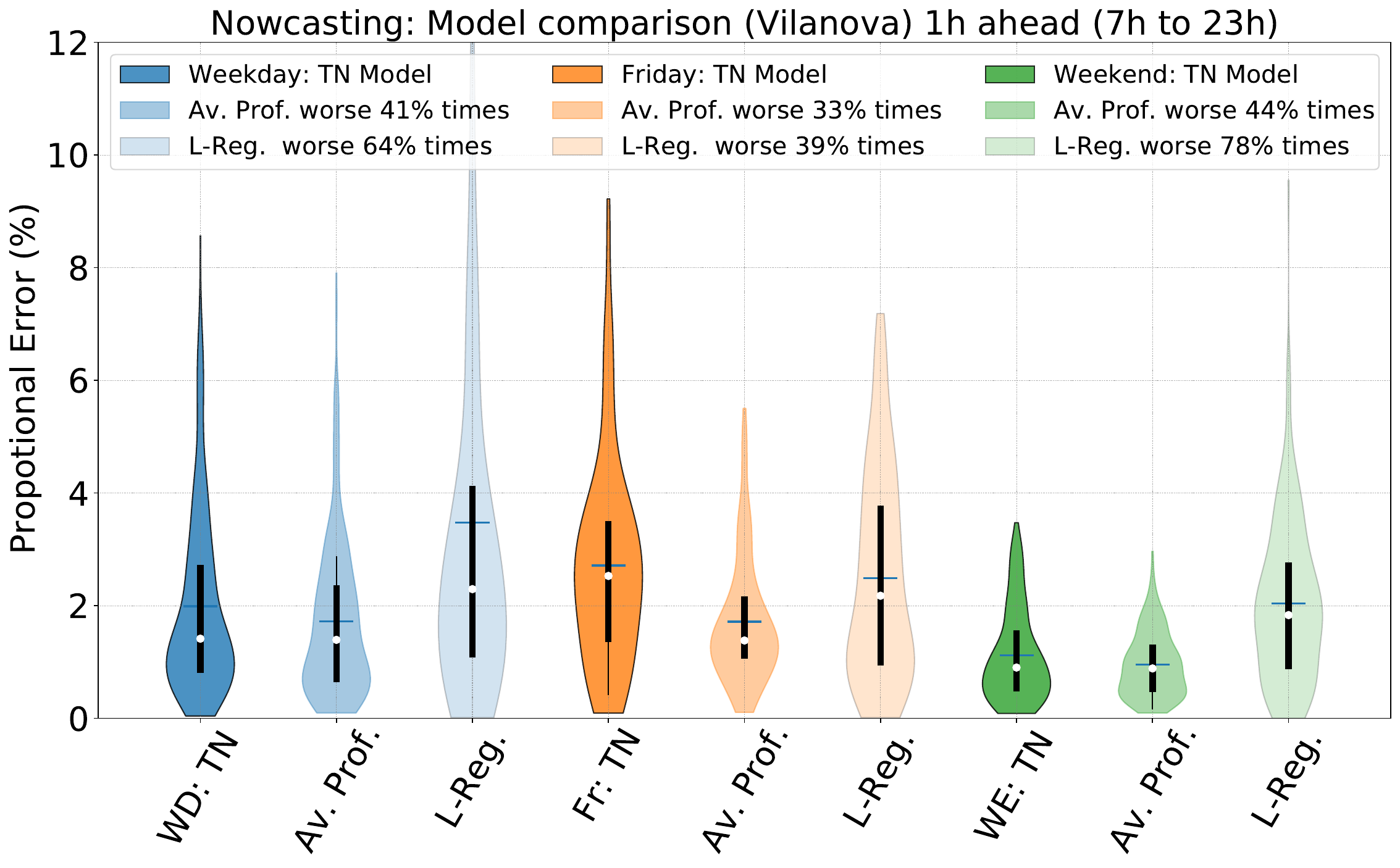}
\includegraphics[width=.45\textwidth]{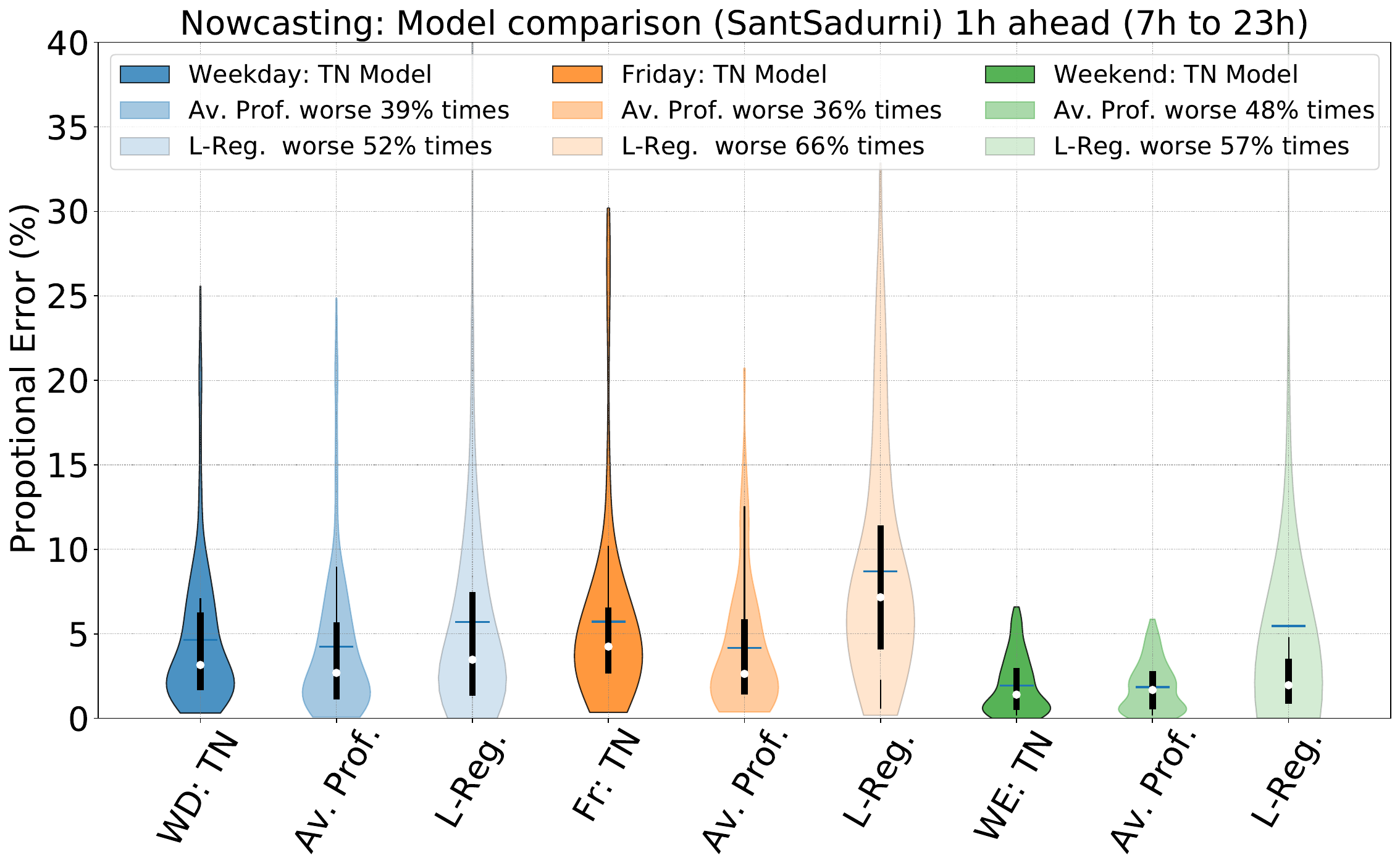}
\includegraphics[width=.45\textwidth]{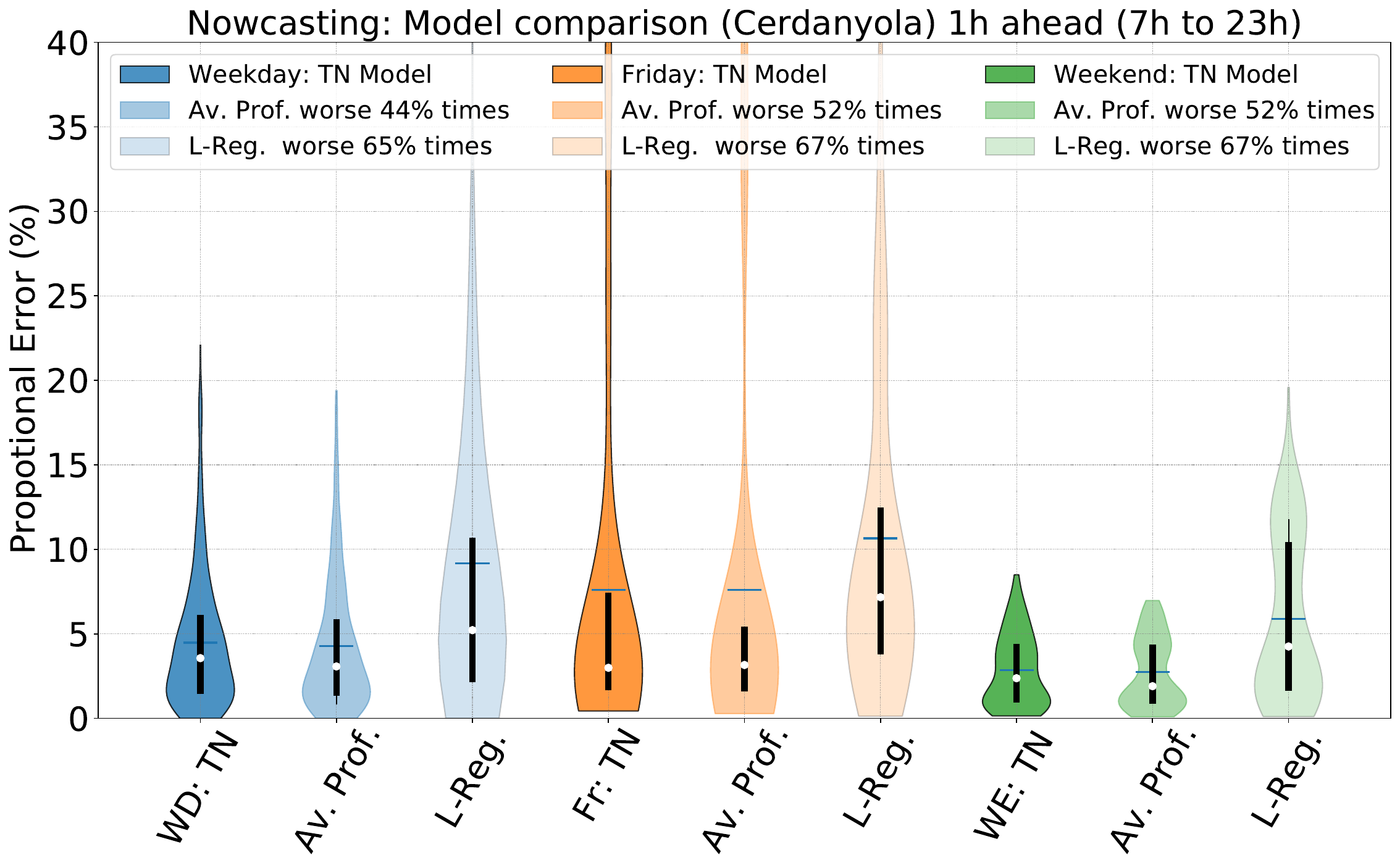}
\includegraphics[width=.45\textwidth]{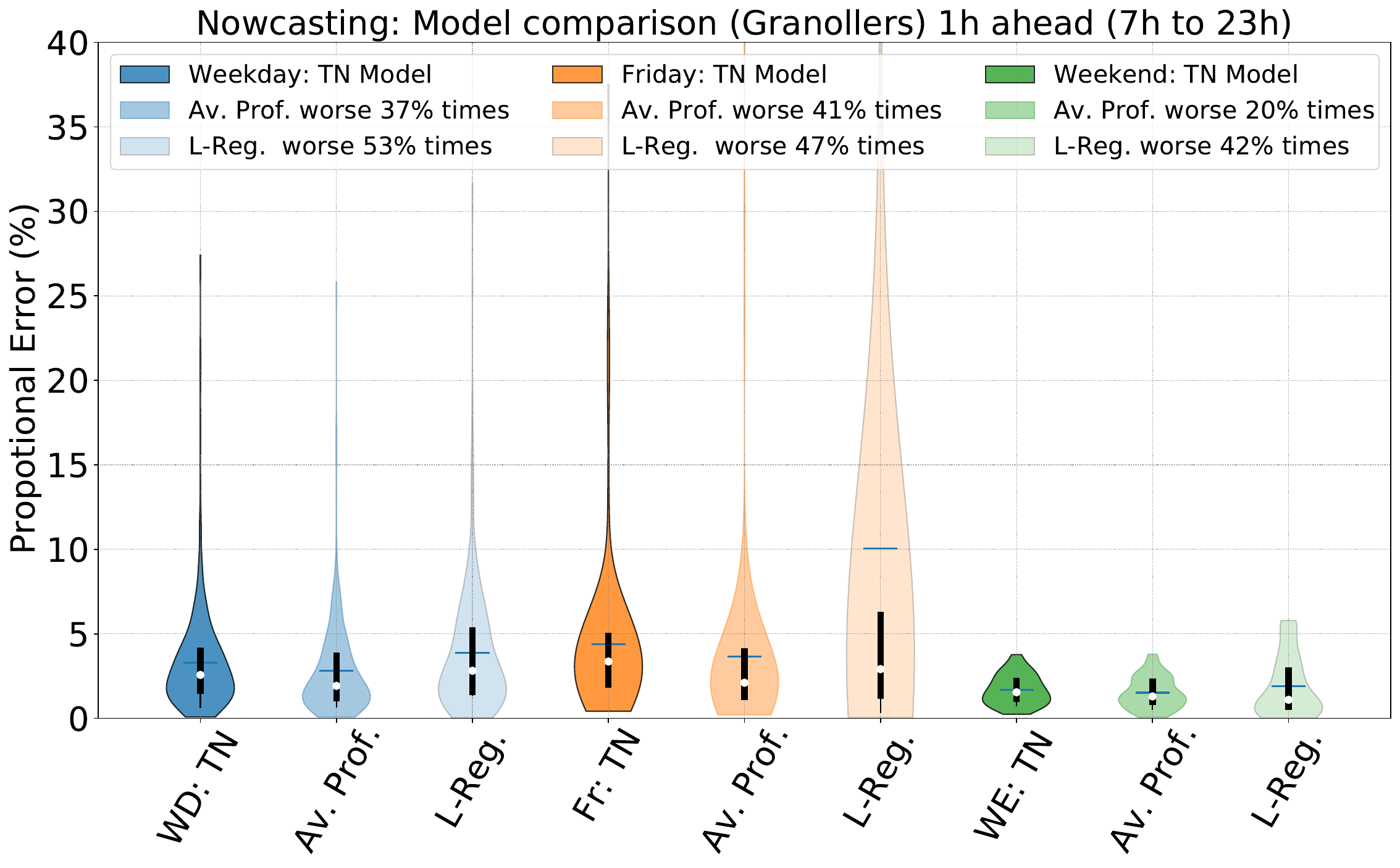}
\includegraphics[width=.45\textwidth]{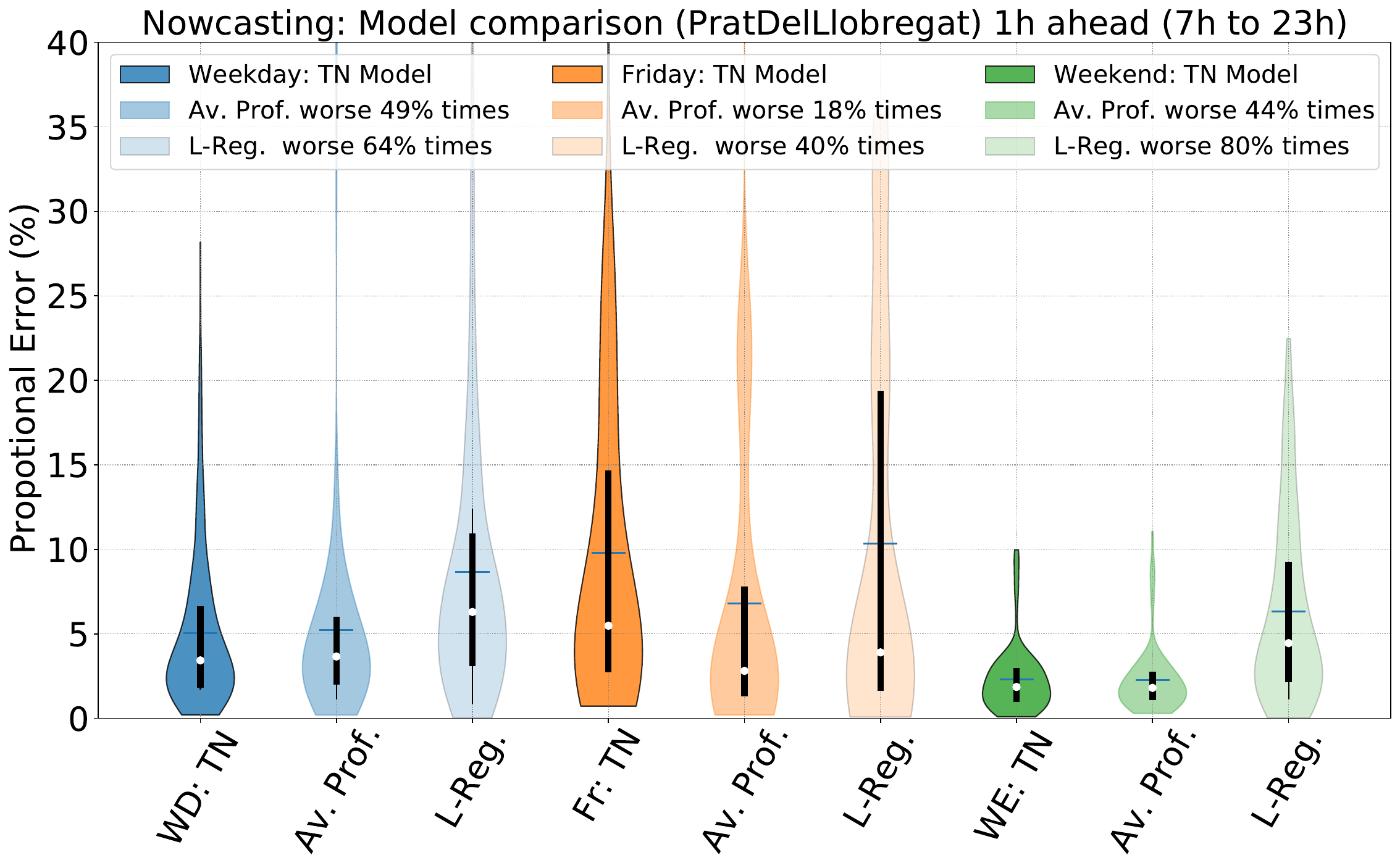}
\includegraphics[width=.45\textwidth]{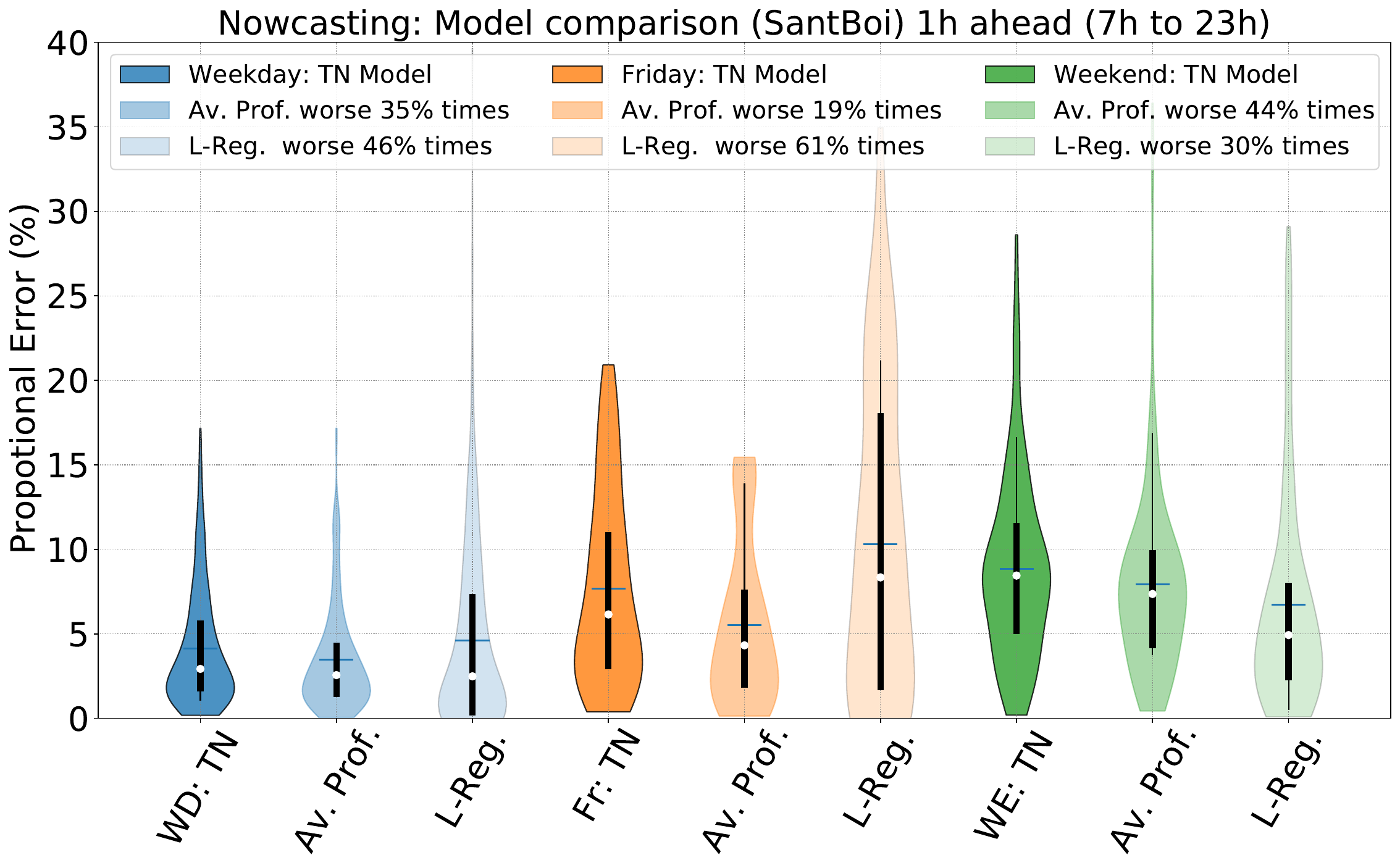}
\includegraphics[width=.45\textwidth]{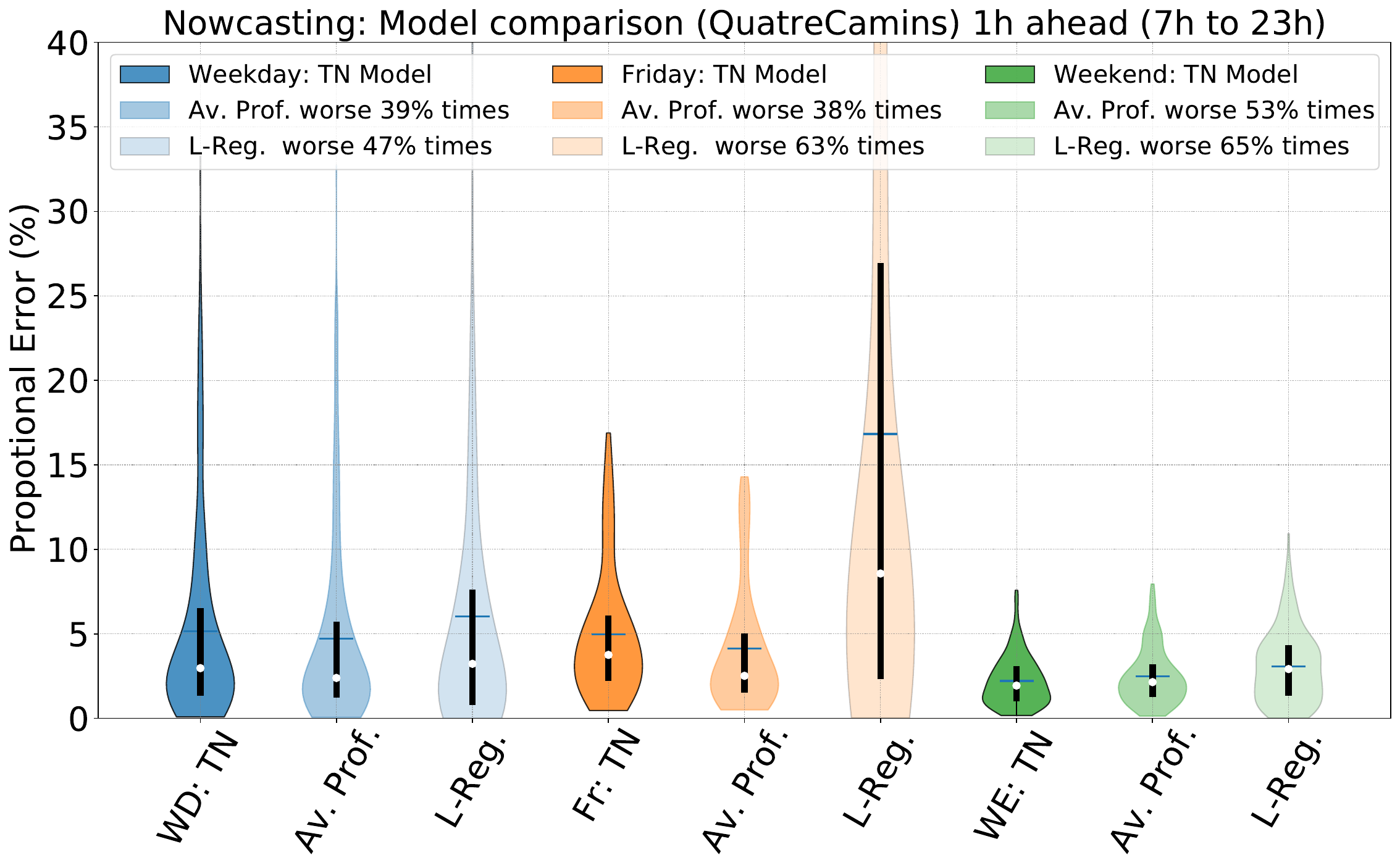}
\includegraphics[width=.45\textwidth]{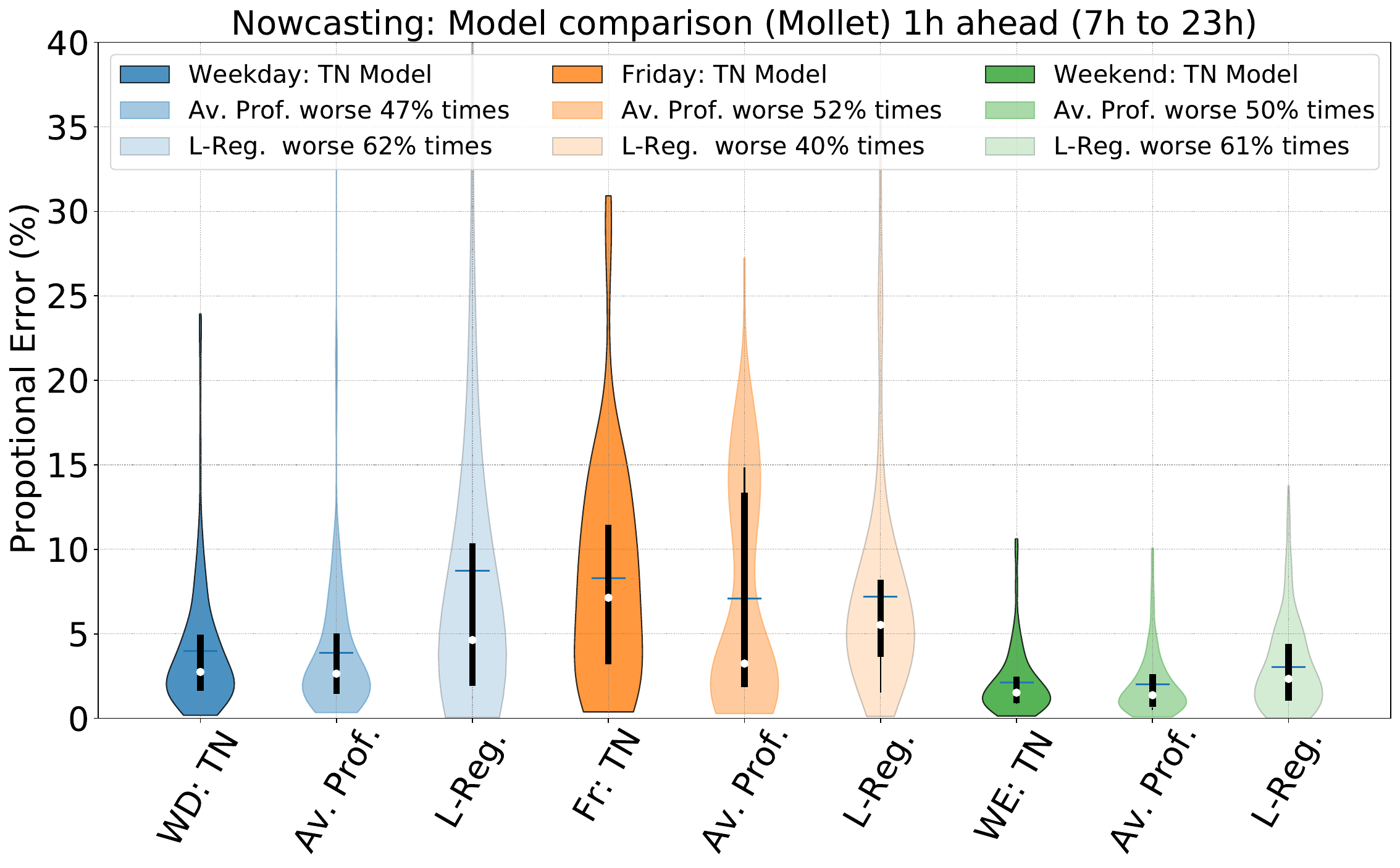}
\caption{
Violin plots of the relative nowcasting error (\%) for different parking lots, predicting the number of cars in the parking area during the next hour based on data available up to a given time. Darker violins represent the performance of the \textbf{TN model}, while brighter violins show comparisons with an \emph{average day-cycle profile}. The brightest violin plots correspond to a simple \emph{linear regression}. Nowcasting is performed from 7:00 to 23:00. Blue horizontal lines indicate the mean of the distributions, the white dot represents the median, and the black vertical bar spans the 25th to 75th percentiles. The legend includes a pairwise performance comparison with the TN model.}
\label{fig:NowCastViolins}
\end{figure}

Figure~\ref{fig:NowCastViolins} shows violin plots of the distribution of nowcasting errors, aggregated by car park and activity profile. The figure compares prediction errors between the basic model (darker colours) and the average activity profile (brighter colours).  

Both approaches exhibit similar performance, with the average profile performing slightly better on average. As before, performance is worst on Fridays (orange tones), improves on weekdays from Monday to Thursday (blue tones), and is best on weekends (green tones), although the latter is likely caused by the lower number of parked cars on weekends in general.

Nowcasting is performed from 7:00 to 23:00 in 30-minute intervals, with predictions made one hour ahead.

\clearpage

\begin{figure}[ht]
\centering
\includegraphics[width=.45\textwidth]{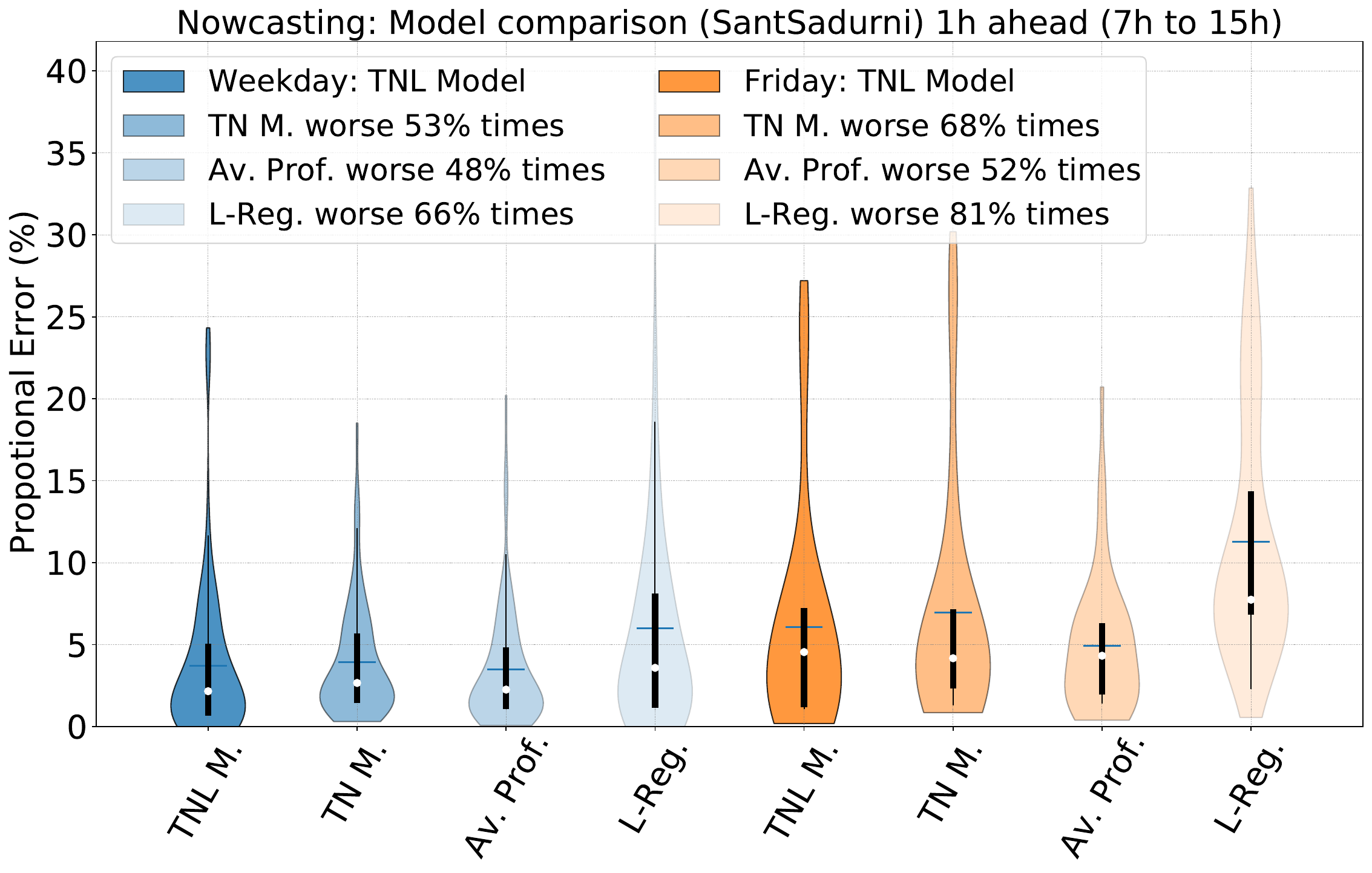}
\includegraphics[width=.45\textwidth]{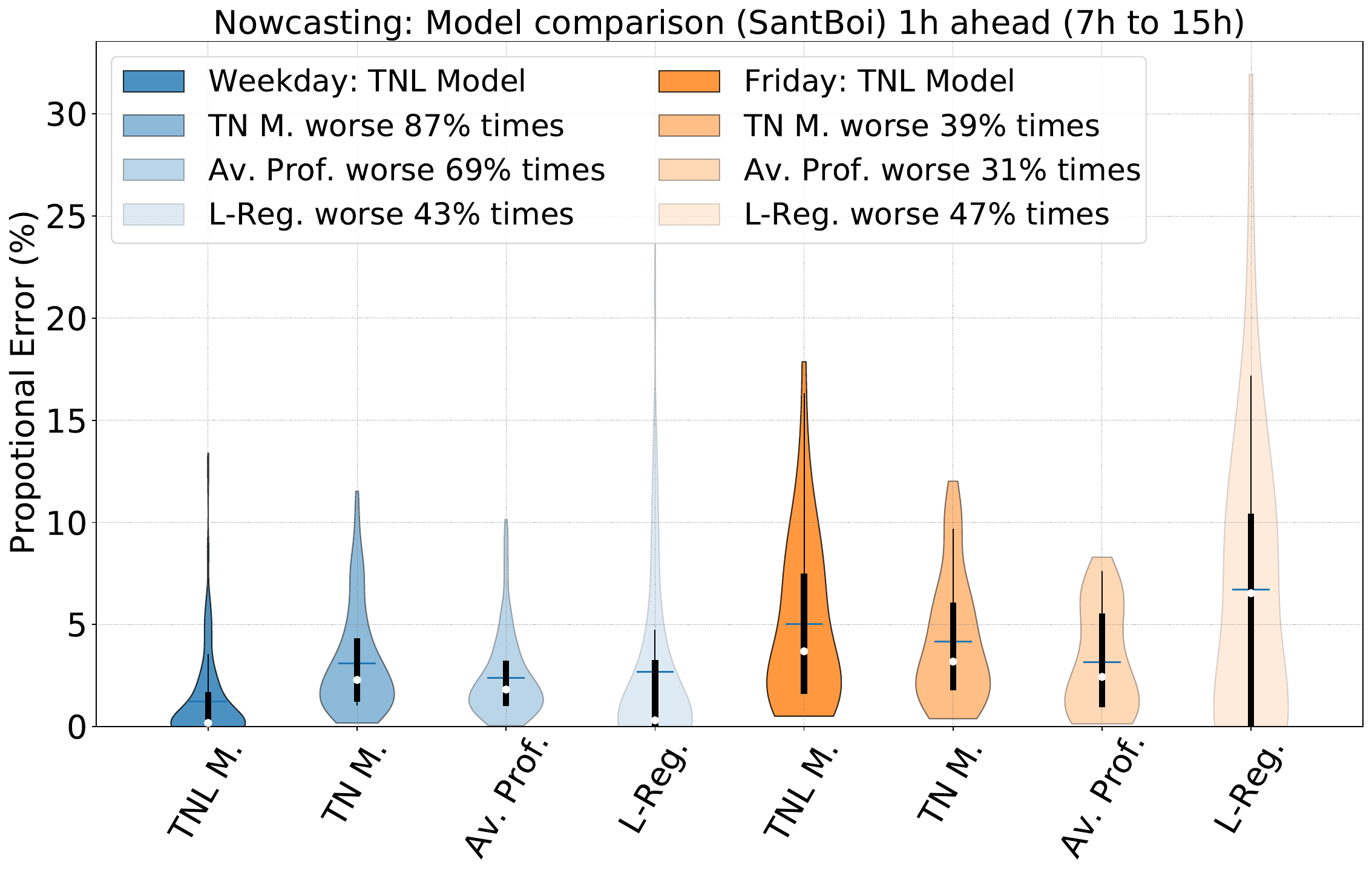}
\includegraphics[width=.45\textwidth]{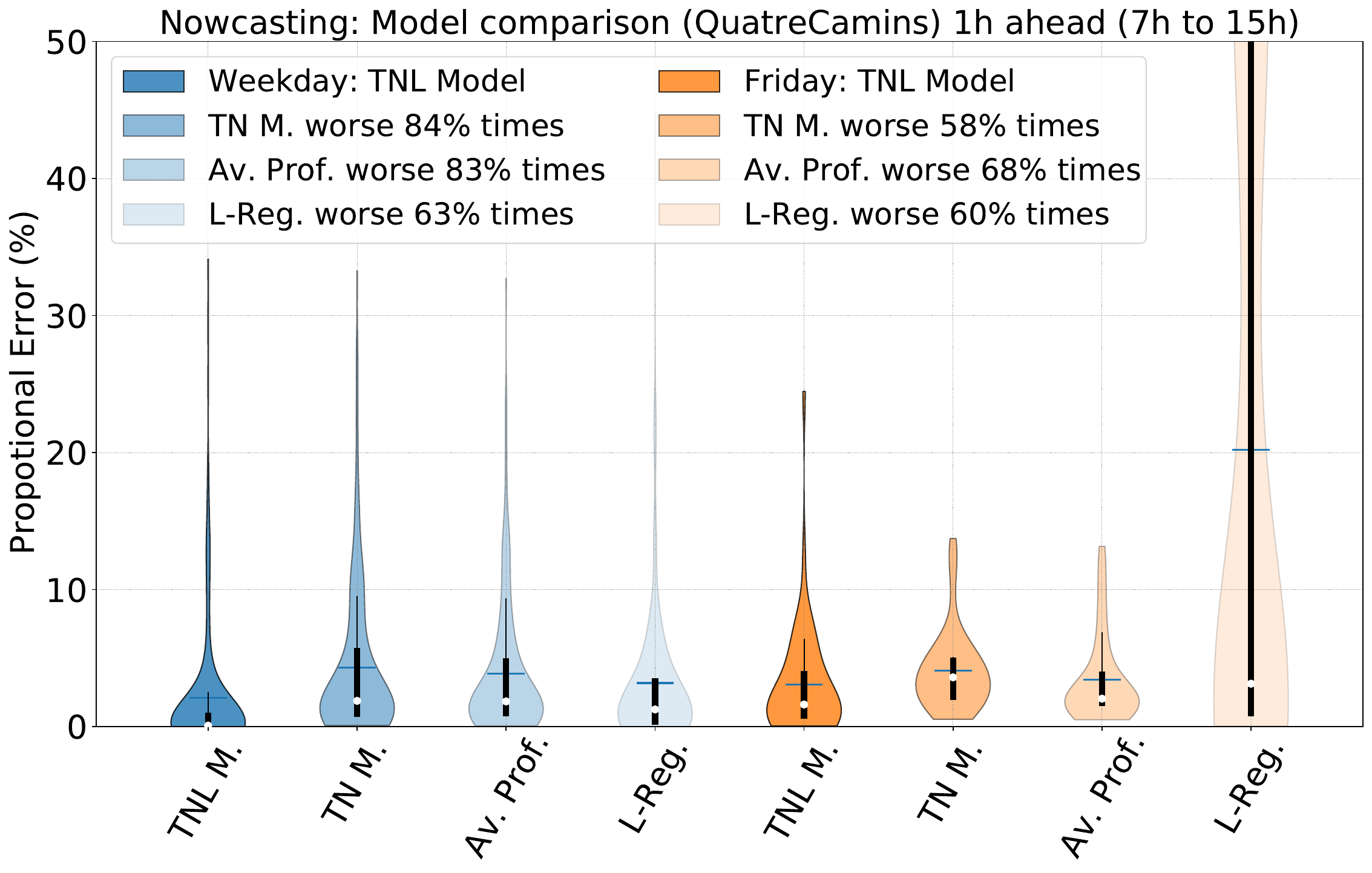}
\includegraphics[width=.45\textwidth]{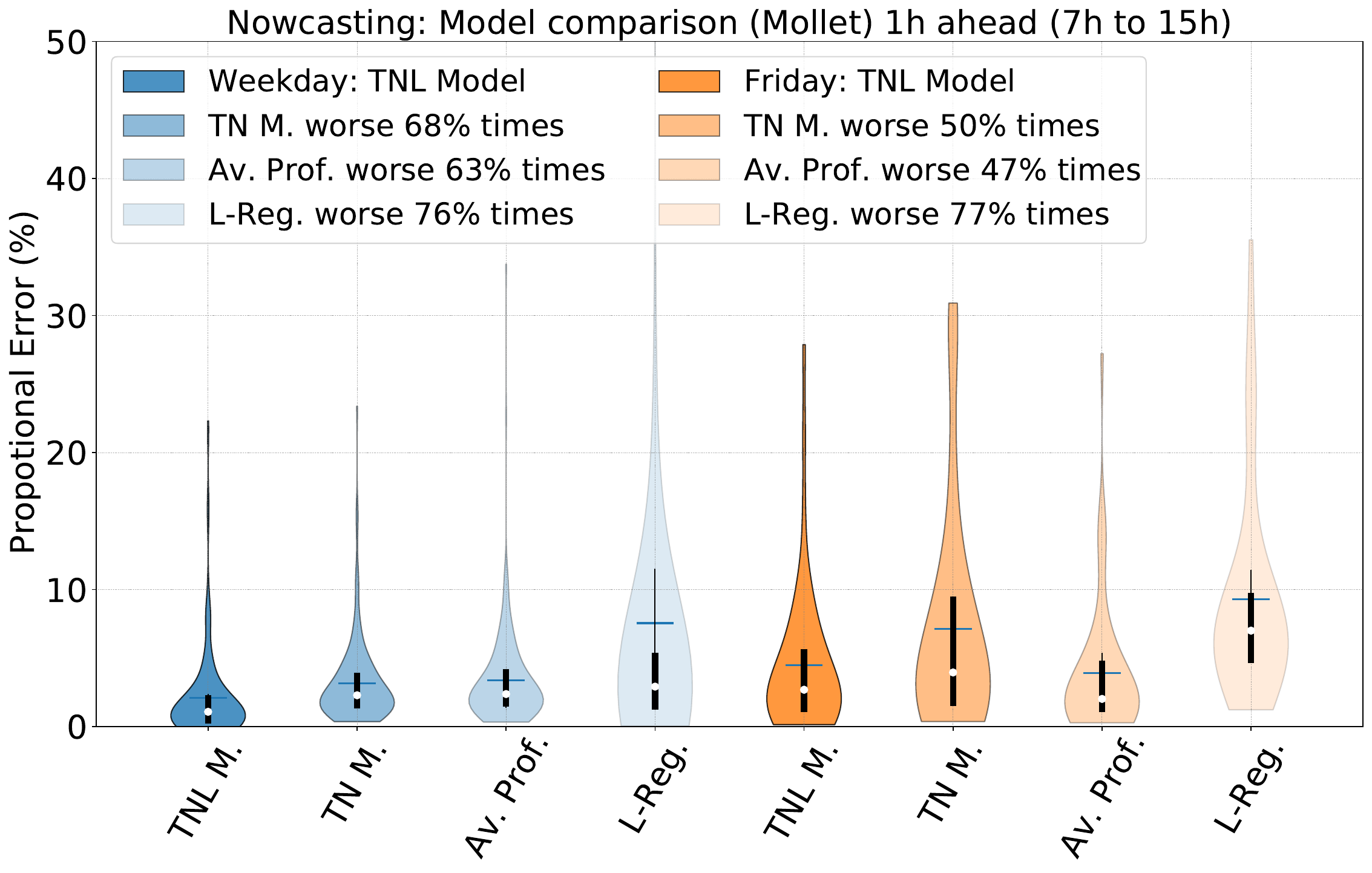}
\caption{
Violin plots of the relative nowcasting error (\%) for different parking lots, predicting the number of cars in the parking area during the next hour based on data available up to a given time. The darkest violins represent the performance of the \textbf{TNL model} with thresholds, slightly brighter violins correspond to the \textbf{baseline TN model}, the second brightest violins represent the \emph{average day-cycle profile}, and the brightest violins correspond to the \emph{Linear Regression model}.  
Nowcasting is performed from 7:00 to 15:00. Blue horizontal lines indicate the mean of the distributions, the white dot represents the median, and the black vertical bar spans the 25th to 75th percentiles. The legend includes a pairwise performance comparison with the TNL model.}
\label{fig:NowCastViolinsTh}
\end{figure}

Figure~\ref{fig:NowCastViolinsTh} compares the performance of the model with activity thresholds for the four stations that reach their capacity limit. In this case, we display nowcasting errors only for the period between 7:00 and 15:00 (with 30-minute intervals). After 15:00, car parks begin to empty, reducing the advantage of using the threshold-based model.  

Violin plots with intermediate colour tones represent the models with thresholds, which consistently outperform the other models on weekdays across all three stations. However, the improvement on Fridays is less pronounced, with only a slight advantage observed for the Quatre Camins car park.  

The figures also include a pairwise comparison of prediction improvement (shown in the legend), where we count how often the TNL model outperforms the other two models in each nowcasting instance. For example, on weekdays at the Quatre Camins station, the TNL model performs better than the baseline model in \textbf{81\%} of cases and surpasses the average profile model in \textbf{79\%} of the tested instances.

For better comparison, Table~\ref{tab:NowCastMedinaTH} presents the numerical values of the corresponding median prediction errors.  

We observe that the Linear Regression Model generally exhibits the highest error values, particularly on Fridays, suggesting that it may not effectively capture key behavioural patterns. In contrast, the TNL Model demonstrates superior performance in predicting parking occupancy, especially on weekdays, where it consistently yields lower error values than the other models.  

Notably, it performs exceptionally well for the Sant Boi and Mollet car parks, achieving median prediction errors as low as \textbf{0.18\%} and \textbf{0.08\%}, respectively, possibly due to more regular commuter patterns.

\begin{table}[!t]
\caption{Median nowcasting error in \% for the four stations depicted in Figure~\ref{fig:NowCastViolinsTh} for weekdays and Fridays comparing the Truncated Normal with Limit (TNL), basic (TN), and Average Profile (Av. Prof.) model with a Linear Regression (L-Reg.). Nowcasting is performed between 7:00 and 15:00 in 30-minute intervals.}
\label{tab:NowCastMedinaTH}
\begin{tabular}{@{}l|rrrr|rrrr@{}}

 \textbf{Car Park} & \multicolumn{4}{c|}{\textbf{weekdays}} & \multicolumn{4}{c}{\textbf{Fridays}}                                                           \\ 
 & TNL & TN  & Av. Prof. & L-Reg. & TNL  & TN  & Av. Prof. & L-Reg. \\ \hline 
 \textbf{Sant Sadurni} & 2.16  & 2.67 & 2.26 & 3.59 & 4.55  & 4.17 & 4.33 & 7.74 \\
\textbf{Sant Boi}     & 0.18 & 2.29 & 1.82 & 0.31 & 3.69 & 3.19 & 2.44 & 6.60 \\
\textbf{Quatre Camins} & 0.08 & 1.88 & 1.84 & 1.26 & 1.62 & 3.60 & 2.04 & 3.14\\
\textbf{Mollet}       & 1.09 & 2.29  & 2.38 & 2.92 & 2.70 & 3.96 & 2.03 & 7.01
\end{tabular}
\end{table} 

The Basic Model, in comparison, generally performs worse than the TNL Model, often showing significantly higher errors. For example, at Sant Boi, its weekday error (2.29) is more than ten times higher than that of the TNL Model (0.18). This highlights the importance of considering capacity limits in the modelling approach for improved accuracy.  

Errors tend to increase across all models on Fridays, suggesting greater variability in parking behaviour. Additionally, since the difference between the TNL and Basic Model becomes less pronounced, this may indicate that, due to lower demand, the capacity limit of car parks is less frequently reached on Fridays in most stations.


\section{Policy Recommendations}
\label{sec:policy}
Based on the findings of this study, we propose the following policy recommendations for transportation planners and municipal authorities involved in managing park-and-ride infrastructure:

\begin{itemize}
    \item {\bf Use simple, explainable models for decision support:}
The proposed Truncated Normal (TN) and Truncated Normal with Limit (TNL) models provide accurate occupancy predictions using a minimal set of interpretable parameters. This simplicity facilitates communication with non-technical stakeholders and supports transparent, evidence-based decision-making.

\item {\bf Monitor and address unmet demand:}
The TNL model explicitly estimates the number of vehicles unable to find parking during peak times. Authorities can use this to identify high-demand stations and prioritise infrastructure investments (e.g., expanding parking capacity or improving alternative access) where overflow is consistently observed.

\item {\bf Incorporate behavioural insights in planning:}
Because the model parameters correspond to actual commuter behaviour (e.g., average arrival and departure times), planners can use these insights to design time-specific interventions (such as staggered work hours, variable pricing, or promotional campaigns encouraging off-peak usage) to better balance demand.

\item {\bf Enable low-cost scalability using aggregate data:}
Our approach requires only aggregate occupancy data, which is often easier and cheaper to collect than individual vehicle tracking. This enables scalable and privacy-preserving deployment across a broad set of car parks without major investments in infrastructure.

\item {\bf Prioritise regularly saturated Facilities:}
As shown in our results (e.g., Figure \ref{fig:occupation}), some car parks consistently reach their capacity limits. We recommend incorporating our model outputs into multi-criteria prioritisation frameworks to guide expansions, especially in facilities critical for multimodal commuting corridors.

\end{itemize}
By aligning parking infrastructure planning with behavioural patterns and unmet demand forecasting, authorities could improve efficiency, user satisfaction, and the overall effectiveness of park-and-ride systems.

\section{Conclusion}

\label{sec:discussion}

We have introduced a simple yet effective model for predicting car park occupancy in park-and-ride facilities. The model leverages truncated normal distributions to represent car arrival and departure times. This approach enables efficient and accurate parking demand prediction and estimation using only a small set of parameters.

A key advantage of the model is that it relies solely on aggregate data, eliminating the need to monitor individual arrivals and departures. Validation using data from Barcelona's metropolitan area demonstrates the model's robustness, even when applied to heterogeneous datasets. Its effectiveness is evident in both prediction and nowcasting tasks.



Furthermore, an extended version of the model, Truncated Normal with Limit (TNL), can estimate the additional parking spaces required to accommodate demand that remains unmet when car parks reach capacity. The extended model explicitly accounts for thresholding behaviour in station overflow, allowing it to determine the moment at which the capacity limit is reached and estimate the number of vehicles unable to find a free parking space.  

This capability is particularly relevant for making informed decisions regarding the expansion of crowded parking lots. This estimation of unmet demand is a particular novelty of our study. To our knowledge, previous research has not directly addressed this issue, despite the fact that parking shortages rank among the top three reasons for not using Park-and-Ride facilities \cite{Zhao2017BehaviorDM}.


Additionally, our work differentiates itself from related studies through its specific focus on the modelling approach. Unlike some studies that prioritise prediction using techniques without an underlying behavioural model~\cite{Zheng2015ParkingAP, Camero2018EvolutionaryDL, zhao2020comparative}, our approach is more ambitious, aiming to understand the fundamental behavioural patterns of commuters rather than merely forecasting occupancy trends.


Moreover, the use of truncated normal distributions provides two key advantages over more complex, overparametrized models, such as neural networks~\cite{Camero2018EvolutionaryDL}. First, it relies on a minimal number of parameters, ensuring computational efficiency. Second, these parameters correspond to behavioural metrics that are inherently interpretable, offering valuable insights into commuter patterns.

In contrast with the prevailing trend of requiring large-scale, fine-grained, high-precision datasets~\cite{xiao2023parking}, our approach demonstrates that accurate predictions can be achieved using aggregated and limited data.

However, we also have to acknowledge two limitations of our study.
First, our model assumes that arrival and departure times are independent random processes. This simplification enables tractability and parameter identifiability using aggregate data. While it may not fully capture dependencies present in individual commuter behaviour (e.g., fixed work shifts), the strong regularity observed in occupancy patterns suggests that, at the aggregate level, this assumption remains a reasonable approximation. Future work could explore coupled distributions or joint modelling frameworks if individual-level data becomes available.
Secondly, the model is tailored to regular commuting behaviour and may be less suitable in non-routine contexts such as shopping centres or event-driven demand, where more flexible or hybrid models may be required.

Nonetheless, and ultimately, we believe this work will contribute to improved urban planning by offering a practical tool for understanding and predicting commuter behaviour in park-and-ride systems. By balancing simplicity, interpretability, and predictive accuracy, our approach can support data-driven decision-making for optimising parking infrastructure and enhancing sustainable mobility solutions.
\backmatter







\section*{List of abbreviations}
\begin{itemize}
    \item \textit{ARIMA}: auto-regressive integrated moving average
    \item \textit{ATM}: Autoritat del Transport Metropolità
    \item \textit{CDF}: cumulative distribution function 
    \item \textit{PDF}: probability density function 
    \item \textit{SVM}: support vector machine
    \item \textit{TN}: Truncated Normal
    \item \textit{TNL}: Truncated Normal with Limit
\end{itemize}

\section*{Declarations}

\subsection{Ethics approval and consent to participate}

 Not applicable.
 
\subsection{Consent for publication}

 Not applicable.

\subsection{Availability of data and materials}
All the data used in the paper and the corresponding code are publicly available at:
\href{https://github.com/aig-upf/car-park}{https://github.com/aig-upf/car-park}

\subsection{Competing interests}
The authors declare that they have no competing interests.

\subsection{Funding}
VG has received funding from “la Caixa” Foundation (ID 100010434), under the agreement LCF/PR/PR16/51110009.

\subsection{Authors' contribution}

AK and VG designed the research project and performed data analysis, model optimisation and evaluation. 
JF and DM performed data analysis, model optimisation and evaluation. 
All authors contributed to the writing and reviewed the manuscript.

\subsection{Acknowledgements}
The authors would like to thank ATM Barcelona (Autoritat del Transport Metropolit\`a) for providing the data.


\bibliography{references}


\begin{thebibliography}{19}
\ifx \bisbn   \undefined \def \bisbn  #1{ISBN #1}\fi
\ifx \binits  \undefined \def \binits#1{#1}\fi
\ifx \bauthor  \undefined \def \bauthor#1{#1}\fi
\ifx \batitle  \undefined \def \batitle#1{#1}\fi
\ifx \bjtitle  \undefined \def \bjtitle#1{#1}\fi
\ifx \bvolume  \undefined \def \bvolume#1{\textbf{#1}}\fi
\ifx \byear  \undefined \def \byear#1{#1}\fi
\ifx \bissue  \undefined \def \bissue#1{#1}\fi
\ifx \bfpage  \undefined \def \bfpage#1{#1}\fi
\ifx \blpage  \undefined \def \blpage #1{#1}\fi
\ifx \burl  \undefined \def \burl#1{\textsf{#1}}\fi
\ifx \doiurl  \undefined \def \doiurl#1{\url{https://doi.org/#1}}\fi
\ifx \betal  \undefined \def \betal{\textit{et al.}}\fi
\ifx \binstitute  \undefined \def \binstitute#1{#1}\fi
\ifx \binstitutionaled  \undefined \def \binstitutionaled#1{#1}\fi
\ifx \bctitle  \undefined \def \bctitle#1{#1}\fi
\ifx \beditor  \undefined \def \beditor#1{#1}\fi
\ifx \bpublisher  \undefined \def \bpublisher#1{#1}\fi
\ifx \bbtitle  \undefined \def \bbtitle#1{#1}\fi
\ifx \bedition  \undefined \def \bedition#1{#1}\fi
\ifx \bseriesno  \undefined \def \bseriesno#1{#1}\fi
\ifx \blocation  \undefined \def \blocation#1{#1}\fi
\ifx \bsertitle  \undefined \def \bsertitle#1{#1}\fi
\ifx \bsnm \undefined \def \bsnm#1{#1}\fi
\ifx \bsuffix \undefined \def \bsuffix#1{#1}\fi
\ifx \bparticle \undefined \def \bparticle#1{#1}\fi
\ifx \barticle \undefined \def \barticle#1{#1}\fi
\bibcommenthead
\ifx \bconfdate \undefined \def \bconfdate #1{#1}\fi
\ifx \botherref \undefined \def \botherref #1{#1}\fi
\ifx \url \undefined \def \url#1{\textsf{#1}}\fi
\ifx \bchapter \undefined \def \bchapter#1{#1}\fi
\ifx \bbook \undefined \def \bbook#1{#1}\fi
\ifx \bcomment \undefined \def \bcomment#1{#1}\fi
\ifx \oauthor \undefined \def \oauthor#1{#1}\fi
\ifx \citeauthoryear \undefined \def \citeauthoryear#1{#1}\fi
\ifx \endbibitem  \undefined \def \endbibitem {}\fi
\ifx \bconflocation  \undefined \def \bconflocation#1{#1}\fi
\ifx \arxivurl  \undefined \def \arxivurl#1{\textsf{#1}}\fi
\csname PreBibitemsHook\endcsname

\bibitem[\protect\citeauthoryear{Lin et~al.}{2017}]{survey}
\begin{barticle}
\bauthor{\bsnm{Lin}, \binits{T.}},
\bauthor{\bsnm{Rivano}, \binits{H.}},
\bauthor{\bsnm{Le~Mouël}, \binits{F.}}:
\batitle{A survey of smart parking solutions}.
\bjtitle{IEEE Transactions on Intelligent Transportation Systems}
\bvolume{18}(\bissue{12}),
\bfpage{3229}--\blpage{3253}
(\byear{2017})
\end{barticle}
\endbibitem

\bibitem[\protect\citeauthoryear{Xiao et~al.}{2023}]{xiao2023parking}
\begin{barticle}
\bauthor{\bsnm{Xiao}, \binits{X.}},
\bauthor{\bsnm{Peng}, \binits{Z.}},
\bauthor{\bsnm{Lin}, \binits{Y.}},
\bauthor{\bsnm{Jin}, \binits{Z.}},
\bauthor{\bsnm{Shao}, \binits{W.}},
\bauthor{\bsnm{Chen}, \binits{R.}},
\bauthor{\bsnm{Cheng}, \binits{N.}},
\bauthor{\bsnm{Mao}, \binits{G.}}:
\batitle{Parking prediction in smart cities: A survey}.
\bjtitle{IEEE Transactions on Intelligent Transportation Systems}
\bvolume{24}(\bissue{10}),
\bfpage{10302}--\blpage{10326}
(\byear{2023})
\end{barticle}
\endbibitem

\bibitem[\protect\citeauthoryear{Zheng et~al.}{2015}]{Zheng2015ParkingAP}
\begin{botherref}
\oauthor{\bsnm{Zheng}, \binits{Y.}},
\oauthor{\bsnm{Rajasegarar}, \binits{S.}},
\oauthor{\bsnm{Leckie}, \binits{C.}}:
Parking availability prediction for sensor-enabled car parks in smart cities.
2015 IEEE Tenth International Conference on Intelligent Sensors, Sensor
  Networks and Information Processing (ISSNIP),
1--6
(2015)
\end{botherref}
\endbibitem

\bibitem[\protect\citeauthoryear{Camero
  et~al.}{2019}]{Camero2018EvolutionaryDL}
\begin{bchapter}
\bauthor{\bsnm{Camero}, \binits{A.}},
\bauthor{\bsnm{Toutouh}, \binits{J.}},
\bauthor{\bsnm{Stolfi}, \binits{D.H.}},
\bauthor{\bsnm{Alba}, \binits{E.}}:
\bctitle{Evolutionary deep learning for car park occupancy prediction in smart
  cities}.
In: \bbtitle{Learning and Intelligent Optimization},
pp. \bfpage{386}--\blpage{401}.
\bpublisher{Springer},
\blocation{Cham}
(\byear{2019})
\end{bchapter}
\endbibitem

\bibitem[\protect\citeauthoryear{Zhao et~al.}{2020}]{zhao2020comparative}
\begin{botherref}
\oauthor{\bsnm{Zhao}, \binits{Z.}},
\oauthor{\bsnm{Zhang}, \binits{Y.}},
\oauthor{\bsnm{Zhang}, \binits{Y.}}:
A comparative study of parking occupancy prediction methods considering parking
  type and parking scale.
Journal of Advanced Transportation,
1--12
(2020)
\end{botherref}
\endbibitem

\bibitem[\protect\citeauthoryear{Awan et~al.}{2020}]{awan2020comparative}
\begin{barticle}
\bauthor{\bsnm{Awan}, \binits{F.M.}},
\bauthor{\bsnm{Saleem}, \binits{Y.}},
\bauthor{\bsnm{Minerva}, \binits{R.}},
\bauthor{\bsnm{Crespi}, \binits{N.}}:
\batitle{A comparative analysis of machine/deep learning models for parking
  space availability prediction}.
\bjtitle{Sensors}
\bvolume{20}(\bissue{1}),
\bfpage{322}
(\byear{2020})
\end{barticle}
\endbibitem

\bibitem[\protect\citeauthoryear{Chen}{2014}]{chen2014parking}
\begin{botherref}
\oauthor{\bsnm{Chen}, \binits{X.}}:
Parking occupancy prediction and pattern analysis.
Dept. Comput. Sci., Stanford Univ., Stanford, CA, USA, Tech. Rep. CS229-2014
(2014)
\end{botherref}
\endbibitem

\bibitem[\protect\citeauthoryear{Ji et~al.}{2015}]{ji2015short}
\begin{barticle}
\bauthor{\bsnm{Ji}, \binits{Y.}},
\bauthor{\bsnm{Tang}, \binits{D.}},
\bauthor{\bsnm{Blythe}, \binits{P.}},
\bauthor{\bsnm{Guo}, \binits{W.}},
\bauthor{\bsnm{Wang}, \binits{W.}}:
\batitle{Short-term forecasting of available parking space using wavelet neural
  network model}.
\bjtitle{IET Intelligent Transport Systems}
\bvolume{9}(\bissue{2}),
\bfpage{202}--\blpage{209}
(\byear{2015})
\end{barticle}
\endbibitem

\bibitem[\protect\citeauthoryear{Chawathe}{2019}]{Chawathe2019UsingHD}
\begin{botherref}
\oauthor{\bsnm{Chawathe}, \binits{S.S.}}:
Using historical data to predict parking occupancy.
2019 IEEE 10th Annual Ubiquitous Computing, Electronics \& Mobile Communication
  Conference (UEMCON),
0534--0540
(2019)
\end{botherref}
\endbibitem

\bibitem[\protect\citeauthoryear{Fokker et~al.}{2022}]{fokker2022short}
\begin{barticle}
\bauthor{\bsnm{Fokker}, \binits{E.S.}},
\bauthor{\bsnm{Koch}, \binits{T.}},
\bauthor{\bsnm{Leeuwen}, \binits{M.}},
\bauthor{\bsnm{Dugundji}, \binits{E.R.}}:
\batitle{Short-term forecasting of off-street parking occupancy}.
\bjtitle{Transportation Research Record}
\bvolume{2676}(\bissue{1}),
\bfpage{637}--\blpage{654}
(\byear{2022})
\end{barticle}
\endbibitem

\bibitem[\protect\citeauthoryear{Vakrinou et~al.}{2025}]{VAKRINOU2025}
\begin{botherref}
\oauthor{\bsnm{Vakrinou}, \binits{K.}},
\oauthor{\bsnm{Mantouka}, \binits{E.G.}},
\oauthor{\bsnm{Vlahogianni}, \binits{E.I.}}:
Leveraging multisource data for off-street parking occupancy prediction: A
  twofold analysis.
International Journal of Transportation Science and Technology
(2025)
\end{botherref}
\endbibitem

\bibitem[\protect\citeauthoryear{Tavafoghi et~al.}{2019}]{Tavafoghi2019AQA}
\begin{botherref}
\oauthor{\bsnm{Tavafoghi}, \binits{H.}},
\oauthor{\bsnm{Poolla}, \binits{K.}},
\oauthor{\bsnm{Varaiya}, \binits{P.}}:
A queuing approach to parking: Modeling, verification, and prediction.
ArXiv
\textbf{abs/1908.11479}
(2019)
\end{botherref}
\endbibitem

\bibitem[\protect\citeauthoryear{Daniotti et~al.}{2023}]{daniotti2023maximum}
\begin{barticle}
\bauthor{\bsnm{Daniotti}, \binits{S.}},
\bauthor{\bsnm{Monechi}, \binits{B.}},
\bauthor{\bsnm{Ubaldi}, \binits{E.}}:
\batitle{A maximum entropy approach for the modelling of car-sharing parking
  dynamics}.
\bjtitle{Scientific Reports}
\bvolume{13}(\bissue{1}),
\bfpage{2993}
(\byear{2023})
\end{barticle}
\endbibitem

\bibitem[\protect\citeauthoryear{Schneble and
  Kauermann}{2025}]{schneble2025statistical}
\begin{botherref}
\oauthor{\bsnm{Schneble}, \binits{M.}},
\oauthor{\bsnm{Kauermann}, \binits{G.}}:
Statistical modelling of on-street parking spot occupancy in smart cities.
Journal of the Royal Statistical Society Series C: Applied Statistics,
017
(2025)
\end{botherref}
\endbibitem

\bibitem[\protect\citeauthoryear{Kaltenbrunner
  et~al.}{2007}]{kaltenbrunner2007description}
\begin{bchapter}
\bauthor{\bsnm{Kaltenbrunner}, \binits{A.}},
\bauthor{\bsnm{G\'omez}, \binits{V.}},
\bauthor{\bsnm{Lopez}, \binits{V.}}:
\bctitle{Description and prediction of slashdot activity}.
In: \bbtitle{Web Conference, 2007. LA-WEB 2007. Latin American},
pp. \bfpage{57}--\blpage{66}
(\byear{2007}).
\bcomment{IEEE}
\end{bchapter}
\endbibitem

\bibitem[\protect\citeauthoryear{Szabo and
  Huberman}{2010}]{szabo2010predicting}
\begin{barticle}
\bauthor{\bsnm{Szabo}, \binits{G.}},
\bauthor{\bsnm{Huberman}, \binits{B.A.}}:
\batitle{Predicting the popularity of online content}.
\bjtitle{Communications of the ACM}
\bvolume{53}(\bissue{8}),
\bfpage{80}--\blpage{88}
(\byear{2010})
\end{barticle}
\endbibitem

\bibitem[\protect\citeauthoryear{Xue et~al.}{2019}]{xue2019commuter}
\begin{botherref}
\oauthor{\bsnm{Xue}, \binits{Y.}},
\oauthor{\bsnm{Fan}, \binits{H.}},
\oauthor{\bsnm{Guan}, \binits{H.}}:
Commuter departure time choice considering parking space shortage and
  commuter’s bounded rationality.
Journal of Advanced Transportation
\textbf{2019}
(2019)
\end{botherref}
\endbibitem

\bibitem[\protect\citeauthoryear{Zhao et~al.}{2017}]{Zhao2017BehaviorDM}
\begin{botherref}
\oauthor{\bsnm{Zhao}, \binits{X.}},
\oauthor{\bsnm{Li}, \binits{Y.}},
\oauthor{\bsnm{Xia}, \binits{H.}}:
Behavior decision model for park-and-ride facilities utilization.
Advances in Mechanical Engineering
\textbf{9}
(2017)
\end{botherref}
\endbibitem

\bibitem[\protect\citeauthoryear{{Generalitat de
  Catalunya}}{1999}]{Gen_strategy}
\begin{botherref}
\oauthor{\bsnm{{Generalitat de Catalunya}}}:
{"PDU for public transport-private vehicle modal interchange car parks within
  the scope of the ATM integrated tariff system in the Barcelona area"}
(1999).
\url{https://shorturl.at/fep5O}
Accessed 5-Mar-2025
\end{botherref}
\endbibitem

\end{thebibliography}

\begin{appendices}

\section{Error Analysis for Two Representative Car Parks}
\label{sec:A1}

In this section, we present the relative errors of the two models for two example car parks.  

\begin{figure}[ht]
\centering
\includegraphics[width=.9\textwidth]{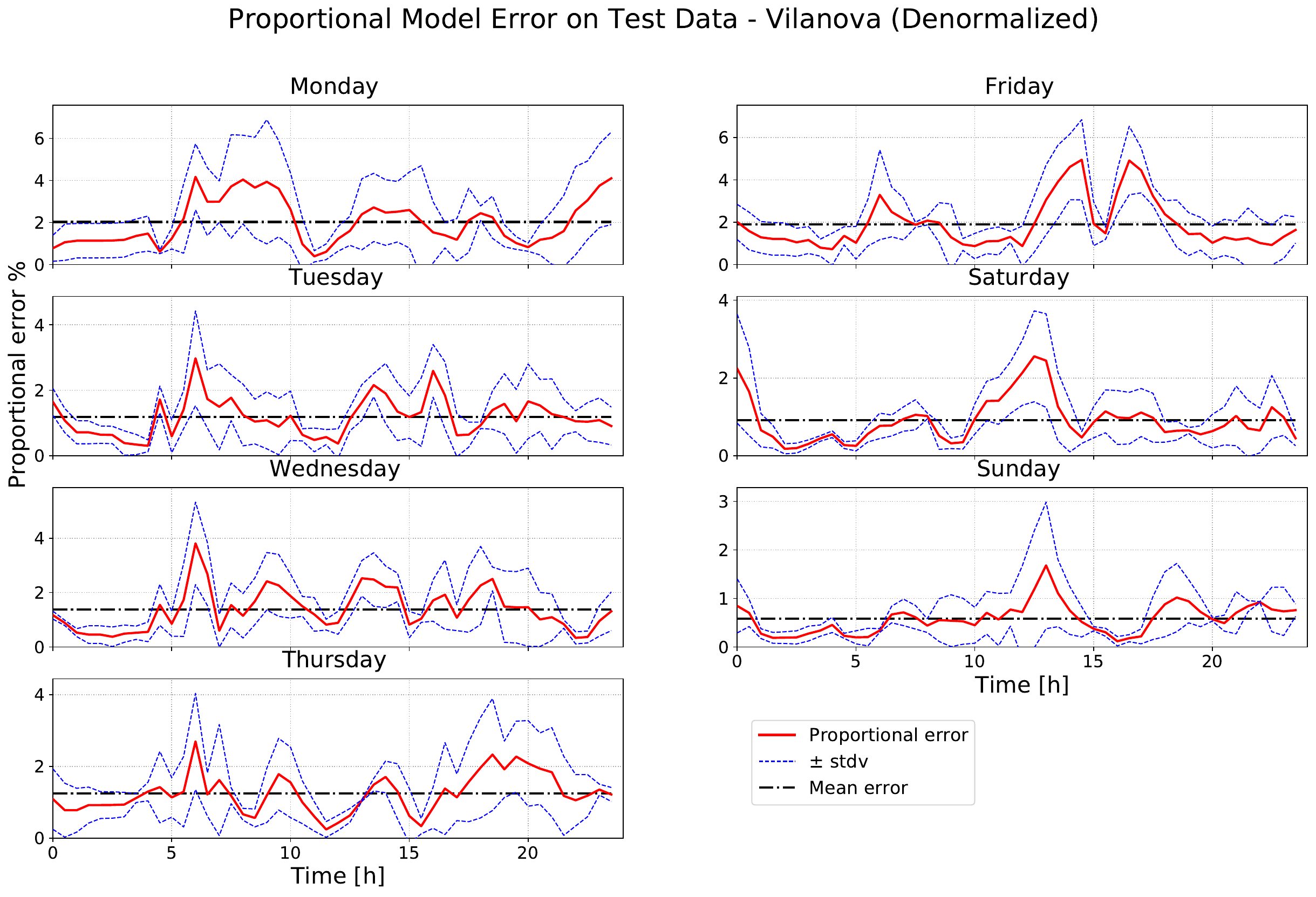}
\caption{{\bf TN Model}: Proportional errors for different weekdays when evaluating the trained model on test data for the Vilanova car park.}
\label{fig:testingErrorVilanova}
\end{figure}

For the TN model, we examine the relative errors depicted in Figure~\ref{fig:testingErrorVilanova} for the Vilanova car park, aggregated by days of the week and hours of the day. The figure displays errors separately for each day of the week, with the average error per hour represented by red continuous lines and the corresponding standard deviations indicated by dashed blue lines. The black dashed-dotted line represents the average error across all hours of the day.  

\begin{figure}[ht]
\centering
\includegraphics[width=.9\textwidth]{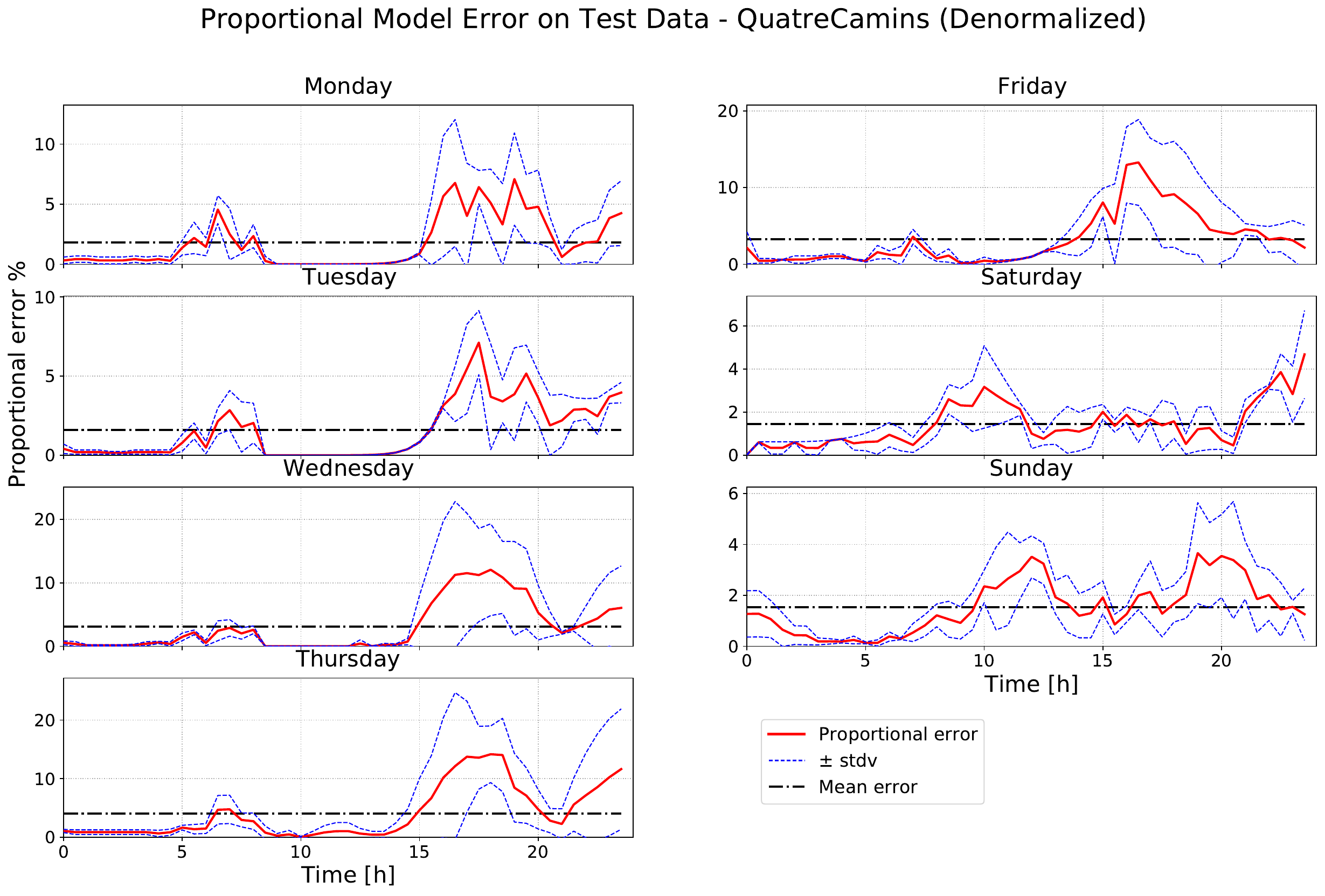}
\caption{{\bf TNL Model}: Proportional errors for different weekdays when evaluating the trained model with an occupancy limit on test data for the Quatre Camins car park. The mean individual training occupancy threshold was used as the model curve.}
\label{fig:testingErrorThQuatreCamins}
\end{figure}

Similarly, Figure~\ref{fig:testingErrorThQuatreCamins} provides an evaluation of the TNL model's performance for the Quatre Camins car park. We observe how this model results in large periods with very small errors but exhibits larger errors in the afternoon during the departure phase of the cars, which is not influenced by the addition of a threshold to the model.

\section{Model parameters and loss}
\label{sec:param}

In this section, we display the values of the parameters obtained when fitting the training data and also the sum of squares loss of the corresponding optimisation processes.

Table~\ref{tab:paramTN} lists the parameters for the TN model for all car parks, and Table~\ref{tab:paramTNL} lists the parameters of the TNL model for the four car parks which reach their capacity limits during weekdays and Fridays.
The corresponding values of the average loss functions per day of training data are given in Tables~\ref{tab:lossTN} and~\ref{tab:lossTNL}. Note that the values of the average loss function are not directly comparable between the TN and the TNL model, as the two models perform different normalisation strategies. 

\begin{table}[!b]
\caption{Parameters $\mu_a\pm\sigma_a$, $\mu_d\pm\sigma_d$    (in hh:mm units) of the TN Model for the different stations. $^\ast$ indicates cases where higher early morning occupancy during weekends (as explained in Section \ref{sec:evaluation}) causes non-interpretable parameter values..}
\label{tab:paramTN}
\begin{tabular}{@{}l|r@{\hspace{1.5mm}}r|r@{\hspace{1.5mm}}r@{}}

\textbf{Car Park} & \multicolumn{2}{c|}{\textbf{Weekdays}} & \multicolumn{2}{c}{\textbf{Fridays}}  \\
  & \multicolumn{1}{c}{arrival} & \multicolumn{1}{c|}{departure}  & \multicolumn{1}{c}{arrival} & \multicolumn{1}{c}{departure}  \\
 \hline 

PratDelLlobregat & 08:07 $\pm$00h58 &  17:21 $\pm$02h50 &  08:36 $\pm$01h25 &  14:22 $\pm$03h48 \\
Granollers & 07:18 $\pm$00h58 &  18:21 $\pm$02h01 &  07:18 $\pm$00h40 &  16:34 $\pm$01h59 \\
Vilanova & 06:56 $\pm$01h16 &  18:40 $\pm$03h05 &  07:02 $\pm$01h35 &  17:27 $\pm$03h33 \\
Mollet & 07:02 $\pm$00h49 &  18:59 $\pm$02h15 &  07:22 $\pm$01h00 &  18:29 $\pm$04h39 \\
QuatreCamins & 07:17 $\pm$00h43 &  19:20 $\pm$01h53 &  07:23 $\pm$00h48 &  18:19 $\pm$02h57 \\
SantSadurni & 07:09 $\pm$01h07 &  19:23 $\pm$02h22 &  07:16 $\pm$02h04 &  18:12 $\pm$03h27 \\
SantBoi & 07:29 $\pm$01h03 &  20:17 $\pm$02h15 &  07:27 $\pm$01h02 &  $^\ast$6809:53 $\pm$164h43 \\
Cerdanyola & 07:11 $\pm$01h19 &  21:33 $\pm$05h46 &  05:52 $\pm$03h23 &  16:42 $\pm$08h27 \\
\hline
\hline
 & \multicolumn{2}{c|}{\textbf{Weekends}} & &  \\
  \hline
PratDelLlobregat & 08:06 $\pm$03h09 &  22:55 $\pm$07h54 & & \\
Granollers & 07:26 $\pm$02h34 &  $^\ast$10786:37 $\pm$249h47 & & \\
Vilanova & 07:50 $\pm$03h00 & $^\ast$248917:52 $\pm$730h15 & & \\
Mollet & 08:34 $\pm$02h46 &  22:39 $\pm$02h13 & & \\
QuatreCamins & 09:29 $\pm$02h08 &  22:53 $\pm$02h31 & & \\
SantSadurni & 09:49 $\pm$01h58 &  19:32 $\pm$04h47 & & \\
SantBoi & $^\ast$00:00 $\pm$07h47 &  $^\ast$33:11 $\pm$02h09 & & \\
Cerdanyola & 08:29 $\pm$04h08 &  18:26 $\pm$07h24 & & \\

\end{tabular}
\end{table}

\begin{table}[!b]
\caption{Parameters $\mu_a\pm\sigma_a$, $\mu_d\pm\sigma_d$    (in hh:mm units) and $\tau$ (in \%) of the TNL Model for the different stations which reach the occupancy limits. $^\ast$ indicates a case where higher early morning occupancy during weekends (as explained in Section \ref{sec:evaluation}) causes non-interpretable parameter values.}
\label{tab:paramTNL}
\begin{tabular}{@{}l|r@{\hspace{2mm}}r@{\hspace{2mm}}r|r@{\hspace{1.5mm}}r@{\hspace{1.5mm}}r@{}}
 \textbf{Car Park} & \multicolumn{3}{c|}{\textbf{Weekdays}} & \multicolumn{3}{c}{\textbf{Fridays}}  \\
 & \multicolumn{1}{c}{arrival} & \multicolumn{1}{c}{departure} & avg. $\tau$ & \multicolumn{1}{c}{arrival} & \multicolumn{1}{c}{departure} & avg. $\tau$\\
 \hline
Mollet & 07:06 $\pm$00h52 & 19:00 $\pm$02h16 & 80.66\% &  07:20 $\pm$00h56 & 19:28 $\pm$04h07 & 72.48\%  \\
QuatreCamins & 07:32 $\pm$00h52 & 19:25 $\pm$01h51 & 79.58\% &  07:43 $\pm$00h55 & 18:30 $\pm$02h50 & 73.08\%  \\
SantSadurni & 07:20 $\pm$01h17 & 19:20 $\pm$02h30 & 76.90\% &  07:08 $\pm$01h47 & 18:17 $\pm$02h57 & 66.69\%  \\
SantBoi & 08:00 $\pm$01h22 & 20:19 $\pm$02h07 & 75.75\% &  11:05 $\pm$02h13 & $^\ast$32:07 $\pm$07h27 & $^\ast$12.29\%  \\
\end{tabular}
\end{table}

\begin{table}[!ht]
\caption{Average Loss per day of the TN Model for the different stations.}
\label{tab:lossTN}
\begin{tabular}{@{}l|ccc@{}}
\textbf{Car Park} & \textbf{Weekdays} & \textbf{Fridays} & \textbf{Weekends}\\
\hline
PratDelLlobregat & 0.00179 & 0.00638 & 0.00541 \\
Granollers & 0.00029 & 0.00246 & 0.00346 \\
Vilanova & 0.00014 & 0.00042 & 0.00174 \\
Mollet & 0.00058 & 0.00269 & 0.00215 \\
QuatreCamins & 0.00026 & 0.00031 & 0.00342 \\
SantSadurni & 0.00021 & 0.00089 & 0.00589 \\
SantBoi & 0.00025 & 0.00057 & 0.01242 \\
Cerdanyola & 0.00209 & 0.00474 & 0.01082 \\
\end{tabular}
\end{table}

\begin{table}[!t]
\caption{Average Loss per day of the TNL Model for the different stations.}
\label{tab:lossTNL}
\begin{tabular}{@{}l|cc@{}}
\textbf{Car Park} & \textbf{Weekdays} & \textbf{Fridays} \\
\hline
Mollet & 0.331 & 1.072 \\
QuatreCamins & 0.141 & 0.149 \\
SantSadurni & 0.135 & 0.459 \\
SantBoi & 0.159 & 0.395 \\
\end{tabular}
\end{table}

\end{appendices}
\clearpage

\section*{Figure Legends}

\listoffigures

\section*{Table Legends}

\listoftables

\end{document}